\documentclass[fleqn,usenatbib]{mnras}

\usepackage{newtxtext,newtxmath}

\usepackage[T1]{fontenc}

\DeclareRobustCommand{\VAN}[3]{#2}
\let\VANthebibliography\thebibliography
\def\thebibliography{\DeclareRobustCommand{\VAN}[3]{##3}\VANthebibliography}

\usepackage{graphicx}	
\usepackage{amsmath}	

\usepackage{enumitem}
\setlist[enumerate]{font={\bfseries}}
\setlist[enumerate,1]{label={(\arabic*)},labelwidth=*}

\defcitealias{Laporte18}{L18}

\title[Disc's vertical response to an infalling satellite]{Measuring the vertical response of the Galactic disc to an infalling satellite}
\author[E. Poggio et al.]{Eloisa Poggio$^{1,2},$\thanks{E-mail: eloisa.poggio@inaf.it}
Chervin F. P. Laporte$^{3,4,2}$, 
Kathryn V. Johnston$^{5,6}$, 
Elena D'Onghia$^{7}$, 
\newauthor
Ronald Drimmel$^{1}$ 
and Douglas Grion Filho$^{5}$ 
\\
$^{1}$Osservatorio Astrofisico di Torino, Istituto Nazionale di Astrofisica (INAF), I-10025 Pino Torinese, Italy\\
$^{2}$Université Côte d’Azur, Observatoire de la Côte d’Azur, CNRS, Laboratoire Lagrange, France\\
$^{3}$ Kavli Institute for the Physics and Mathematics of the Universe (WPI), The University of Tokyo Institutes for Advanced Study (UTIAS),\\
The University of Tokyo, Chiba 277-8583, Japan\\
$^{4}$ Department of Physics \& Astronomy, University of Victoria, 3800 Finnerty Road, Victoria BC, V8P 5C2 Canada\\
$^{5}$Department of Astronomy, Columbia University, 550 West 120th Street, New York, NY 10027, USA \\
$^{6}$Center for Computational Astrophysics, Flatiron Institute, 162 5th Av., New York City, NY 10010, USA\\
$^{7}$University of Wisconsin, Madison, Astronomy Department, 475 N Charter Str., Wisconsin, USA
}

\date{Accepted XXX. Received YYY; in original form ZZZ}

\pubyear{2015}

\begin{document}
\label{firstpage}
\pagerange{\pageref{firstpage}--\pageref{lastpage}}
\maketitle

\begin{abstract}
%
Using N-body simulations of the Milky Way interacting with a satellite similar to the Sagittarius dwarf galaxy, we quantitatively analyse the vertical response of the Galactic disc to the satellite's repeated impacts. 
We approximate the vertical distortion of the Galactic disc as the sum of the first three Fourier azimuthal terms $m = 0, 1$ and $2$, and observe their evolution in different dynamical regimes of interaction. After the first interaction, the $m=0$ term manifests itself as outgoing ring-like vertical distortions. The $m=1$ term (S-shape warp) is prograde when the impacts of the satellite are more frequent, or in general close to an interaction, whereas it is slowly retrograde in the most quiescent phases.  The $m=2$ term is typically prograde, and close to an interaction it couples with the $m=1$ term. 
Finally, we find that the vertical response of the disc can be recovered in an unbiased way using the instantaneous positions and velocities of stars in a limited volume of the Galactic disc, analogous to real data, and that the measured vertical pattern speeds have a constraining power in the context of a Milky Way-satellite interaction. 

\end{abstract}










\section{Introduction}

The Milky Way's vertical structure may hold important clues of its past formation. As soon as 21-cm surveys made it possible to study the distribution of neutral hydrogen gas around edge-on galaxies, it became apparent that many galaxies, including our own, are warped \citep{Burke57,Kerr57}. With the advent of large photometric surveys, such as the 2MASS and Sloan Digital Sky Survey, it became evident that the disc exhibited more complex behaviour in its vertical structure. In the solar neighbourhood, \cite{Widrow12} uncovered signs of vertical waves in the form of North/South asymmetries in number density and bulk velocity of stars. The kinetic signatures of these waves were confirmed in \citet{williams13,carlin13, schoenrich18}. Moreover, a growing number of observational studies suggested these waves could be part of much larger scale ones, giving rise to a warped and rippled disc \citep{xu15,Price-Whelan15,li17}.

A number of mechanisms have been proposed to explain the existence/persistence of warps in galaxies. These have ranged from special configurations (and assumptions) such as the presence of flattened potential around discs \citep{Sparke88,Dubinski95}, torques exerted by rotating halos \citep{Nelson95}, misaligned angular momenta between halo and disc \citep{debatista99} to misaligned gas infall \citep{LopezCorredoira02} and the influence of external perturbers \citep{Weinberg98}, in particular through the excitation of global dark matter wakes which can penetrate deep into the Galactic midplane as well as tidal interactions. In particular satellite interactions are a prime mechanism to producing not only warps \citep[e.g.][]{Gomez17,Semczuk20}, but also the more complex structure such as the generation of bending waves \citep[e.g.][]{Gomez13, widrow14, Donghia16b}, rings \citep{younger08, kazantzidis09} and kinematic features as tidal tails \citep{toomre72, DOnghia2010, Laporte19a} 
which are seen in the Milky Way. In fact, tidal encounters are central to the lifetime of galaxies in a cosmological context, as every central galaxy is surrounded by a subhalo population which can in some cases (depending on the mass and distance reached by a subhalo) shape the evolution of Galactic discs \citep[e.g.][Giammaria et al., submitted]{font01, gauthier06,aumer13}. Given the characteristic functional form of the subhalo mass function \citep{springel08}, the less numerous but much more dynamically impactful subhalos are the more massive ones.

In the Milky Way, potential suspects have long been the Magellanic Clouds \citep{Weinberg06} and the Sagittarius dwarf \citep{ibata98,Bailin03} galaxy (though with mixed success until more recently). While the Magellanic Clouds are more massive, they are also more distant, and new proper motion measurements \citep{kallivayalil13} indicate that they are on a first infall, implying only a very recent interaction. However, first infall N-body models of the Large Magellanic Cloud \citep{laporte18a} show that they are still able to warp the Galactic disc similarly to the HI observations \citep{Levine06} through the excitation of a wake which may be traced using the stellar halo \citep[e.g.][]{Garavito19}. On the other hand, Sagittarius (hereafter, Sgr) is currently undergoing full disruption, and has been 
perturbing the Galaxy for a more extended period of time through numerous pericentric passages. Moreover, recent estimates of its total stellar progenitor mass place it at higher infall halo masses \citep[e.g.][]{gibbons17} which significantly raises its potential to shape the Galaxy. In particular, it has been proposed that Sgr could be an architect for 
spiral structure in the Milky Way \citep{purcell11} and the formation of bending waves seen in the solar neighbourhood and at large distances \citep{Gomez13} at least in a qualitative sense. Recent N-body simulations of Sgr-like interactions with the Milky Way \citep{Laporte18} show that it is possible to simultaneously reproduce the amplitude of the density fluctuations in the solar neighbourhood and the extent of known outer disc structures \citep{Newberg02,Rocha-Pinto04}.

Thanks to the second \emph{Gaia} Data Release \citep[hereafter, \emph{Gaia} DR2, ][]{2018A&A...616A...1G}, the complexity in kinematics and richness in terms of substructure of the Galactic disc has been mapped to greater detail and unprecedented spatial coverage around the Sun \citep{2018A&A...616A..11G}. Vertical disturbances, streaming motions, wave patterns and arches in the velocity space 
have been revealed on a large scale \citep{Ramos18,Poggio18,RomeroGomez19,Friske19}, triggering new interest and questions into the physical mechanisms regulating the dynamical evolution of the disc \citep{Carrillo19, 2020A&A...634A..66L, 2020NatAs...4..590P}. Complementary to the earlier findings of \citet{Widrow12}, when viewed in 2D, the vertical phase-space distributions of stars in the solar neighbourhood revealed spiral snail-shell features \citep{Antoja18}, signaling that the disc is out of equilibrium and currently undergoing phase-mixing, possibly caused by Sgr \citep{Antoja18, BS18, Laporte18, Bland-Hawthorn19, Haines2019, Laporte19c, Bland-Hawthorn2020}. Alternatively, phase-space spirals have been suggested as a manifestation of internal mechanisms, such as the bar buckling instability \citep{Khoperskov19}. The sharper view of Gaia DR1 combined with the SDSS's long baseline to provide accurate proper motions, and later those of \emph{Gaia} DR2, has also revealed remnants and confirmed earlier signs of ancient massive accretion events, among which stands out the Gaia Enceladus-Sausage galaxy \citep{meza05,navarro11,Belokurov18,Helmi18,Haywood18,RecioBlanco18,Helmi20}, thought to have occurred about 10 Gyr ago.

With the increasing complexity faced by the data and the number of competing dynamical processes which could occur in concert, it is now imperative to study and develop theoretical models to help with the interpretation of observations. Ultimately, one would want to discriminate between the different dynamical processes that might be at play (bar buckling, misaligned gas accretion, halo rotation, satellite(s) interaction(s)...). Recently, the vertical response of galactic discs to multiple external perturbers has been revisited in some detail by \cite{2018MNRAS.480.4244C}, who compared the behaviour of bending waves in the presence of a population of dark matter subhalos and in isolation. However, the tidal effects are dominated by the most massive and luminous substructures, and many subhalos never reach the central part of a halo \citep{font01}. Thus, a complementary approach to \cite{2018MNRAS.480.4244C} is to study the response of a Galactic disc subject to a minor merger over a lifespan of several orbital timescales on a polar orbit. This should help get a better intuition of the effect of a particular kind of orbital configuration shared by at least one massive known luminous satellite of the MW, the currently disrupting Sgr dwarf.

In this work we aim to explore the reaction of a Milky Way-like disc to a massive Sgr-like satellite as it spirals deeper into the galactic potential. The forthcoming Gaia Data Release 3 and other future datasets will afford even more detailed and global views of the Galaxy, motivating us to perform a Fourier analysis of the disc vertical distortion, which may be feasible with stellar data. In this paper, we aim to: (i) characterize the response of the disc to the repeated impacts of the dwarf galaxy, (ii) study the connection between the satellite's orbit and the disc response; (iii) investigate whether the response might be reliably measured from a limited data volume; (iv) understand whether the vertical response of the disc has a constraining power in the context of a Milky Way-Sgr-like interaction. In a companion work, we look at the influences that drive this evolution (Filho et al., in prep.).

The paper is structured as follows. In Section \ref{Sec:Sims}, we describe the simulations used in this study. In Section \ref{Sec:Regimes}, we define and discuss different dynamical regimes present in the MW-Sgr interaction process. Section \ref{Sec:VertResp} is dedicated to the analysis of the disc vertical response to the repeated interactions with the Sgr-like satellite. In Section \ref{Sec:obs_disc_response}, we compare the actual evolution of the disc obtained in Section \ref{Sec:VertResp} to the results that would be obtained from a volume-limited survey. Finally, we summarise and discuss our our results in Section \ref{Sec:summary}.



\begin{figure*}
	\includegraphics[width=14cm]{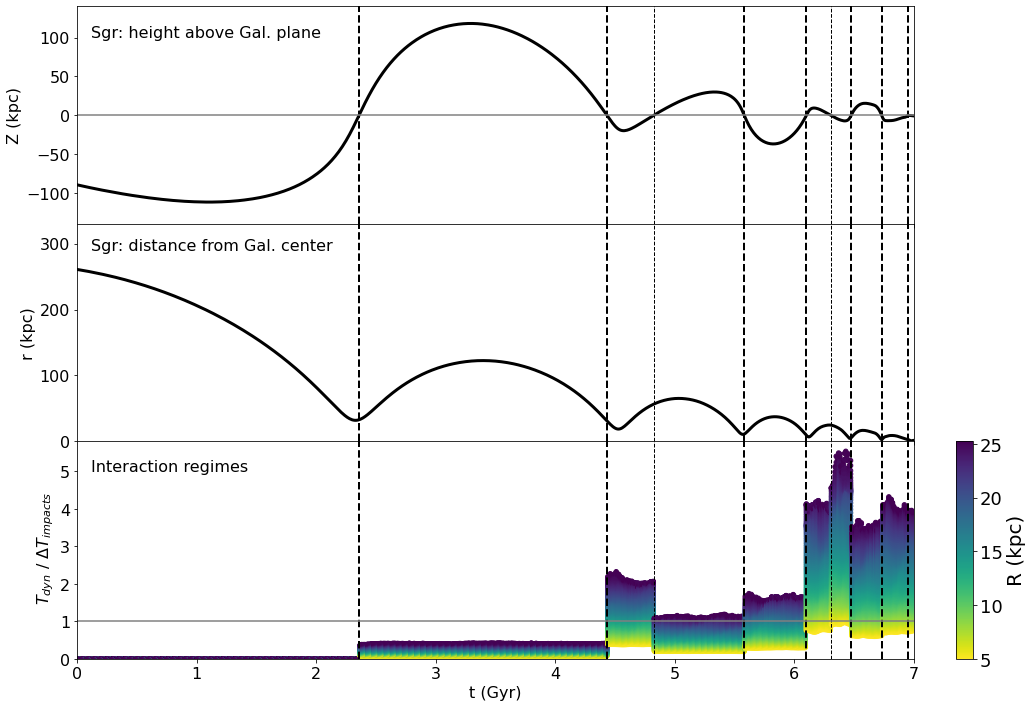}
    \caption{ \emph{Upper panel}: Vertical elevation of Sgr with respect to the Galactic plane as a function of time, based on Sgr L2 model. \emph{Middle panel}: Same as before, but showing the distance from the Galactic center. \emph{Lower panel}: Regimes of dynamical interaction between Sgr and the Milky Way, defined as described in Section \ref{Sec:Regimes}. The grey horizontal line shows where $ T_{\rm{dyn}} /  \Delta T_{\rm{impacts}} = 1$, which roughly corresponds Regime 2. In all three panels, the dashed vertical lines indicate Sgr's plane-crossings (i.e. where Z=0). Due to a different response of the disc (see text), we have made a distinction between the plane-crossings in proximity (thick dashed vertical lines) and distant from (thin dashed vertical lines) Sgr's nearest pericenter.
    \label{fig:regimes}}
\end{figure*}

\begin{figure*}
	\includegraphics[width=17cm]{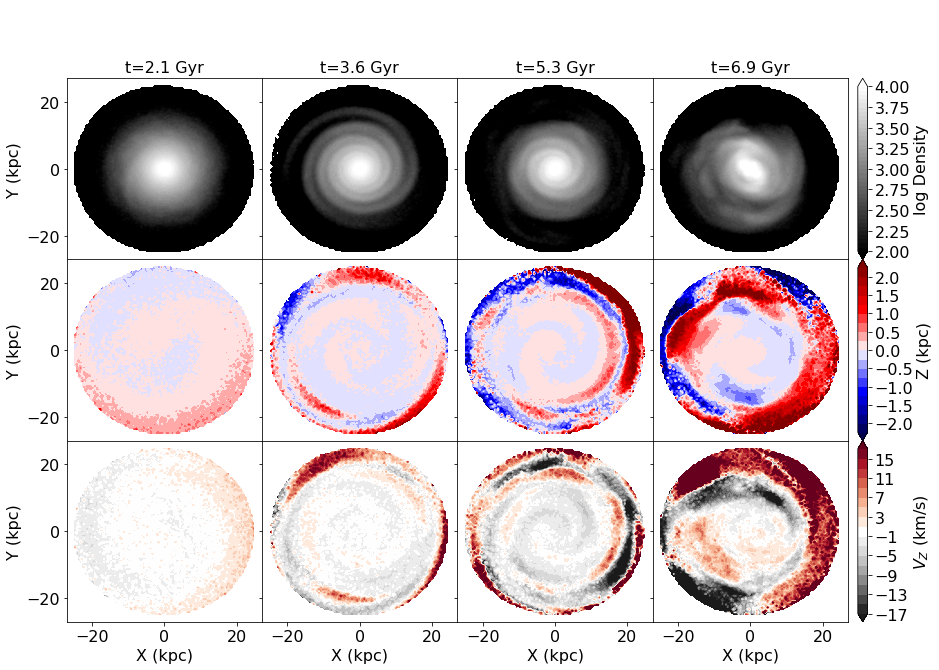}
    \caption{Face-on views of the Milky Way's disc in the four interaction regimes (see text: surface density (upper row), mean vertical coordinate Z (middle row) and vertical velocity Vz (lower row) at different times.  At t=2.1 Gyr (first column) no impacts with Sgr have occured. At t=3.6 Gyr (second column), t=5.3 Gyr (third column), t=6.9 Gyr (fourth column), the galactic disc has experienced, respectively, one, three and eight impacts. Disc rotation is clockwise. 
    }
    \label{fig:Z_Vz_maps}
\end{figure*}


\section{Simulations}  \label{Sec:Sims}


The simulations analysed here are taken from the pre-Gaia DR2 runs performed in \cite{Laporte18} \citepalias[hereafter][]{Laporte18}. These consist of a series of controlled N-body experiments aimed at studying the response of a MW-like system to Sgr and LMC-like perturbers. Despite its modest setup, this simulation was used to establish that the large scale perturbations in the outer disc such as the Monoceros Ring complex \citep{Newberg02, Slater14, Deason18}, TriAnd \citep{Rocha-Pinto04} and small scale inner disc perturbations \citep{Widrow12, Antoja18} can be simultaneously understood as the response of the Galactic disc to perturbations from a satellite like Sgr \citep{Laporte18, Laporte19a}. A number of observaitonal studies had suggested a disc interpretation for outer stellar halo-disc interface structures \citep{Price-Whelan15, xu15, li17, deBoer18, Sheffield18} and were later confirmed through chemistry \citep[e.g.][]{Bergemann18,Hayes18}. It was also the first N-body simulation used to interpret the Gaia DR2 RVS velocity fields and phase-space spiral in Gaia DR2 and study its formation with a realistic model of a Sgr-like satellite interaction \citep{Laporte19c}. The simulations were also used to predict the formation of structures like the Anticenter stream \citep{Laporte19a} as remnants of tidal tails excited by satellite interaction and the expected modulation of the star formation history of the Galaxy following the impacts which was first reported in \citep{Laporte20} using the outer disc Monoceros complex and recently in the solar neighbourhood \citep{Ruiz-Lara2020}. 

In this paper, we take a closer look at the L2 run, which considered the interaction between a MW-like host and a Sgr-like progenitor with $M_{SGR200}=6\times10^{10}\,\rm{M_\odot}$ infalling from the virial radius. For more details on the simulations and their analysis, we refer the reader to \citetalias{Laporte18}, but provide a brief summary here to guide the discussion. For all runs, a fiducial host Milky Way-like model was adopted consisting of an exponential disc in the radial direction and isothermal vertically with scale radius $R_d=3.5\,\rm{kpc}$ and scale height $z_d=0.53\,\rm{kpc}$, a Hernquist bulge with $M_b=10^{10}\,\rm{M_\odot}$ and scale length $a_{b}=0.7\,\rm{kpc}$ surrounded by a dark matter halo of mass $M_{200}=10^{12}\,\rm{M_\odot}$ approximated by a \cite{hernquist90} profile with scale length $a_h=53\,\rm{kpc}$ which is adiabatically contracted according to the prescription of \citet{blumenthal86}. This results in a galaxy with a rotation curve consistent with \cite{mcmillan11}. The initial conditions were realised with {GalIC} \citep{yurin14} and the galaxy was set with a Toomre $Q>1.5$ to avoid the formation of non-axisymmetric features in the isolated case. This choice is arbitrary of course and only intended for the sake of the numerical experiment at the time to understand the response of the disc in the absence of internal perturbations from the start.

Throughout the paper, we analyse the disc's dynamics in the simulation from a reference frame centered around the Galactic Center with respect to the Galactic midplane. This is achieved using the same analysis methods outlined in \cite{Gomez13,Laporte18}, which involve centering the galaxy through the iterative shrinking sphere algorithm \citep{power03} and re-aligning the system with respect to the total angular momentum of the inner disc. As a result, any re-orientation or precession of the disc w.r.t. to its initial configuration resulting from interactions \citep[e.g.][]{velazquez99,Gomez17} will be lost. This dynamical response is not studied in this paper. 


\section{Interaction regimes  } \label{Sec:Regimes}


The interaction process between Sgr and the Galactic disc can be divided into different regimes, based on the the orbit of Sgr and the dynamical properties of the disc. 

At time t=0, the progenitor of Sgr is launched into orbit and crosses the Galactic plane several times before merging fully with the Milky Way. As time passes, plane-crossings become more frequent (Figure~\ref{fig:regimes}, top panel), with Sgr passing through the Galactic plane (z=0 kpc) at 2.36, 4.43, 4.83, 5.58, 6.1, 6.31, 6.48, 6.73 and 6.95 Gyr (dashed vertical lines in Figure Figure~\ref{fig:regimes}). At the plane-crossings, Sgr is located at 32, 30, 55, 11, 9, 24, 5, 2.5 and 1 kpc from the Galactic center, respectively. As can be seen in Figure~\ref{fig:regimes}, most (but not all) of the plane-crossings are close to the orbit pericenters.

As discussed in the following section, the influence of Sgr becomes progressively more important as its orbit gets closer to the MW and the impacts become more frequent. However, the response to Sgr is not the same everywhere in the Galactic disc. 
We define the dynamical time $ T_{\rm{dyn}}$ (i.e. Galactic year) as a function of Galactocentric radius R, i.e. $ T_{\rm{dyn}} \sim 2 \pi R / V_{\phi}$, where $ V_{\phi}$ is the mean azimuthal velocity. For example, for stars rotating at $ V_{\phi} \sim 230$ km/s, the dynamical time would be $\sim $ 220 Myr at $R = 8$ kpc, or $\sim $ 550 Myr at $R = 20$ kpc.

We define different Sgr-MW interaction regimes based on the ratio $ T_{\rm{dyn}} / \Delta T_{\rm{impacts}}$, where $ T_{\rm{dyn}}$ is the above-defined dynamical time of the disc, and $ \Delta T_{\rm{impacts}}$ is the time interval between two impacts of Sgr. As shown in Figure~\ref{fig:regimes} (bottom), the ratio $ T_{\rm{dyn}} / \Delta T_{\rm{impacts}} $ demarcates four different regimes:
\begin{itemize}
\item {\bf Regime 0} or pre-interaction regime: $  T_{\rm{dyn}} / \Delta T_{\rm{impacts}} \sim 0$, as $\Delta T_{\rm{impacts}} \sim \infty$. The disc effectively evolves in isolation.
\item {\bf Regime 1}: $0 <  T_{\rm{dyn}} /  \Delta T_{\rm{impacts}} < 1$. In this Regime, interaction time-scales are long enough to give the disc the possibility to relax between one and the other plane-crossing.
\item {\bf Regime 2}: $ T_{\rm{dyn}} /  \Delta T_{\rm{impacts}} \sim 1$. Since the disc dynamical time is of the same order of the time interval between two impacts, the disc doesn't have time to fully relax after an interaction. 
\item {\bf Regime 3}: $ T_{\rm{dyn}} / \Delta T_{\rm{impacts}} > 1$. This is the most perturbed regime, where the repeated impacts on short time-scales in the inner disc cause a strong and complicated response in the Galactic disc.
\end{itemize}
As shown in Figure~\ref{fig:regimes} (bottom), the interaction regime strongly depends on radius R. As a result, for a given time, different regimes may coexist at different radii.

While this simulation is not an exact reconstruction of the MW-Sgr interaction, it is useful to consider what
point in the simulation is most similar to the present-day Milky Way-Sgr system. Based on current data
\citep{Ibata94,Johnston05,Law10,Penarrubia10}, Sgr's position in the Milky Way is just below the Galactic plane, at about $\sim 16$ kpc from
the Galactic center. Considering the L2 model, the most similar geometry is at the 4$^{\rm{th}}$ and 5$^{\rm{th}}$ plane-crossings (i.e. R$ \sim
10$ kpc), which occurs between Regime 2 and 3.


\begin{figure*}
	\includegraphics[width=12.5cm]{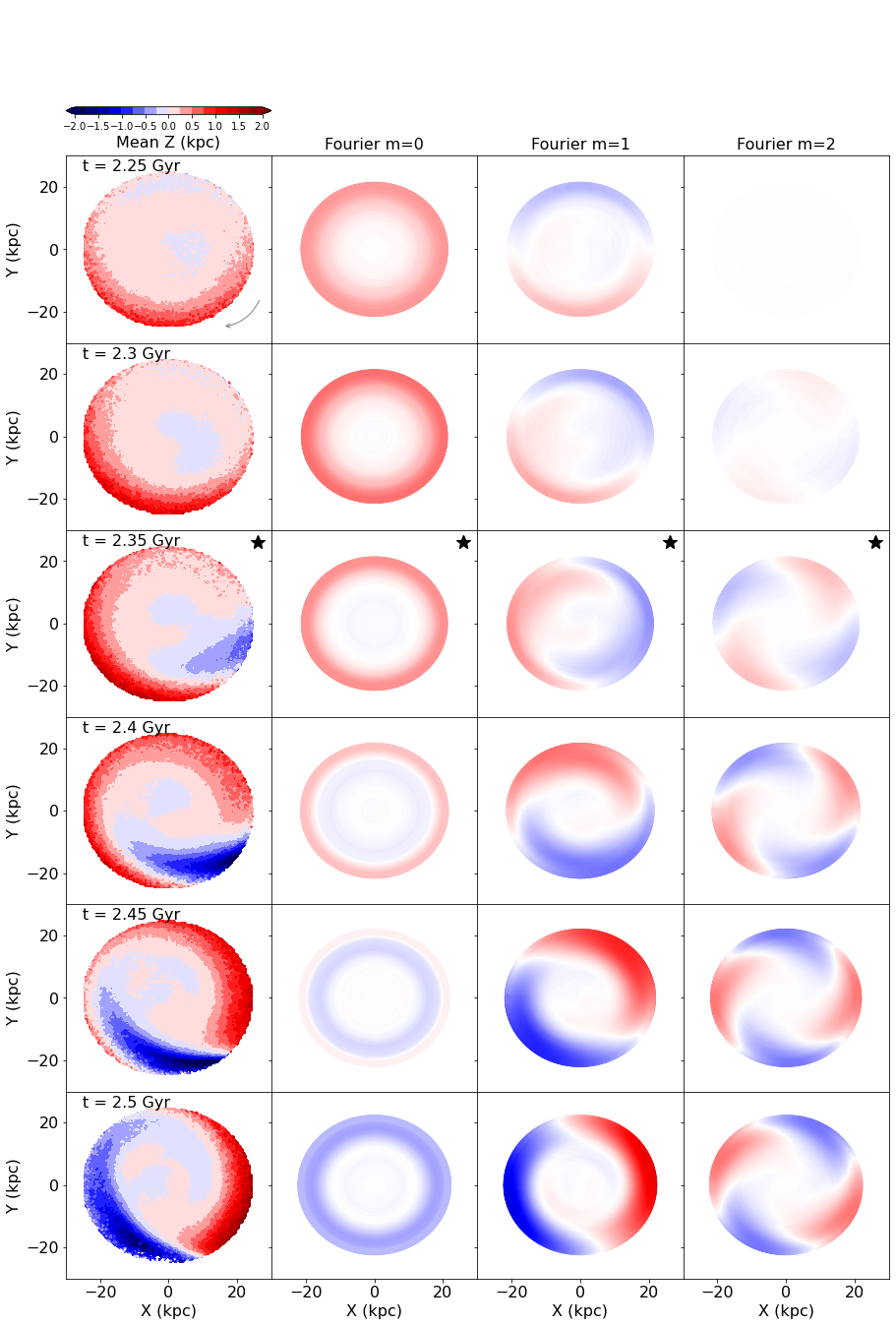}
    \caption{Face-on view of the mean vertical distortion of the Galactic disc during the first interaction with Sgr (first column), accompanied by the corresponding Fourier decomposition into the $m=0$ (second column), $m=1$ (third column) and $m=2$ (fourth column) terms. The Galaxy rotates clockwise. The snapshot at t=2.35 Gyr has been marked by a black star, to highlight that an interaction with Sgr is taking place (the pericenter and plane-crossing occur, respectively, at 2.33 and 2.36 Gyr). The Fourier decomposition shows that: (i) Sgr excites ring-like vertical distortions (second column), which propagate from the inner to the outer parts of the disc; (ii) The $m=1$ distortion moves in a prograde direction with respect to Galactic rotation; (iii) The $m=2$ term is excited by the interaction with Sgr. 
    An animated version of this figure is available for the entire simulation.
    \label{fig:Fourier_decomposition}}
\end{figure*}

\begin{figure*}
	\includegraphics[width=16cm]{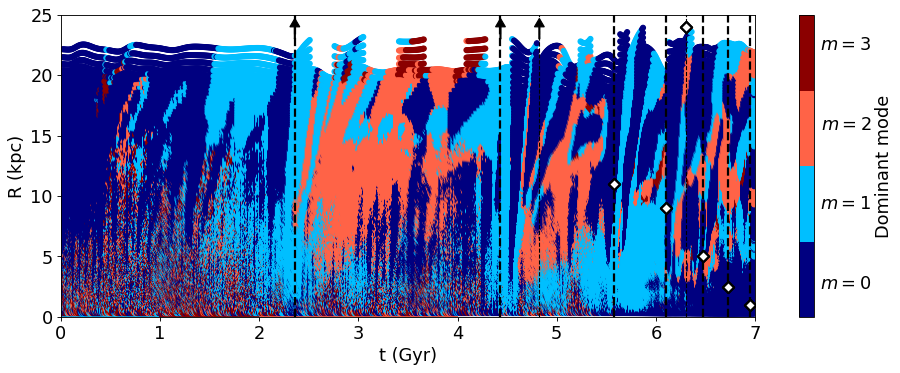}
    \caption{For a given time t and Galactocentric radius R, the Fourier term with the largest amplitude based on our analysis. Dashed vertical lines indicate plane-crossings of Sgr. The position of Sgr during a plane-crossing is indicated by a white diamond if $R_{\rm{Sgr}} < 25$ kpc. Otherwise, a black arrow is drawn at the top of the figure. 
    } 
    \label{fig:amp}
\end{figure*}

\begin{figure*}
	\includegraphics[width=16cm]{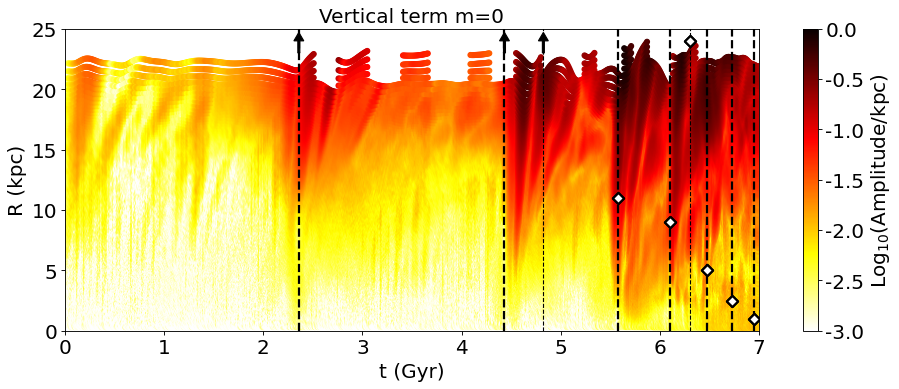}
	\includegraphics[width=16cm]{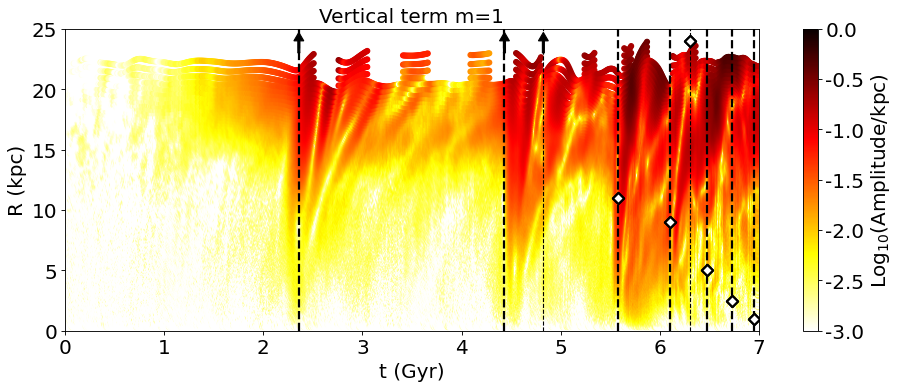}
	\includegraphics[width=16cm]{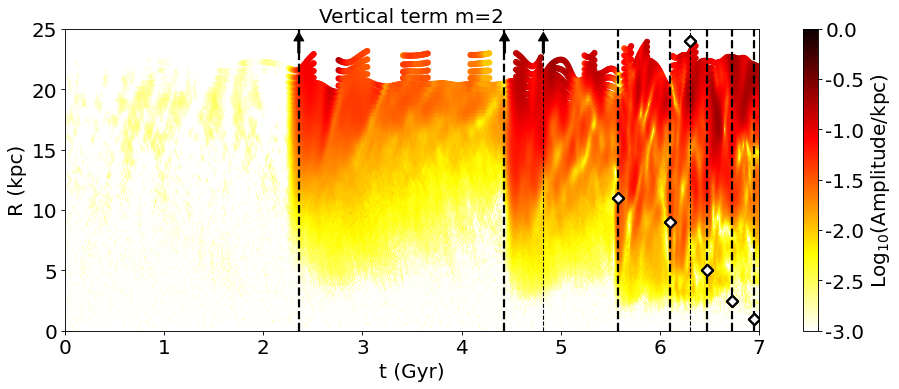}
    \caption{Amplitude of the Fourier vertical terms m=0 (upper panel), m=1 (middle panel)  and m=2 (lower panel). Dashed vertical lines, arrows and white diamonds are as in Figure \ref{fig:amp}.}
    \label{fig:amp_sep}
\end{figure*}

\begin{figure*}
	\includegraphics[width=16cm]{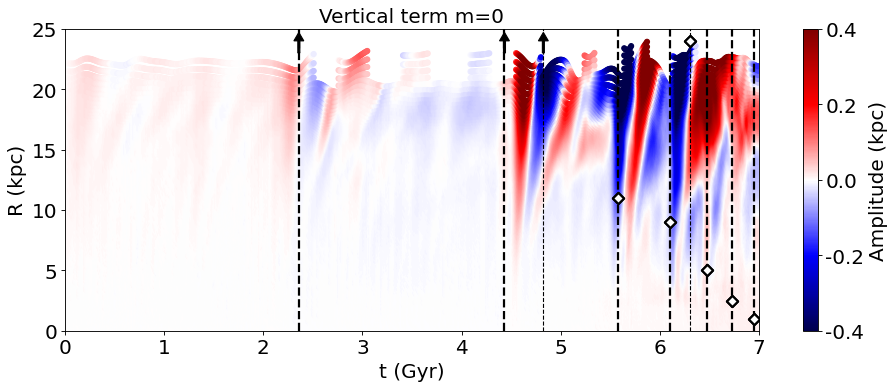}
    \caption{Amplitude of the vertical Fourier m=0 term. Dashed vertical lines, arrows and white diamonds are as in Figure \ref{fig:amp}.
    }
    \label{fig:m0_amp_sign}
\end{figure*}


\section{Vertical response of the disc} \label{Sec:VertResp}

\subsection{Qualitative overview}

Here, we qualitatively discuss the response of the disc to the repeated impacts of Sgr. Although it is not the main objective of this paper, it is instructive to observe the evolution of the disc stellar density (Figure~\ref{fig:Z_Vz_maps}, first row). The disc is initially smooth (first panel), and develops a grand design spiral structure after the first interaction (second panel). After three disc crossings, spiral arms are not as well-defined as before, and present small-scale discontinuities in the outer parts (third panel). Finally, after eight disc crossings, the spiral arms are disrupted and a central bar is formed (fourth panel).

Different interaction regimes clearly manifest themselves also in the vertical dimension (Figure~\ref{fig:Z_Vz_maps}, second row). The disc initially exhibit a mild S-shape warp (first panel), just before the first disc crossing. After the first interaction with Sgr, the distortion of the disc remains relatively smooth, but spiral waves are excited (second panel, see following Sections for a deeper analysis). Such waves can be clearly seen for R < 15 kpc, are trailing, but do not spatially coincide with the spiral arms observed in the density. As plane crossings become more frequent and Sgr gets closer to the disc, the vertical distortion becomes more prominent (third panel), until vertical features loose their coherence over large portions of the Galactic disc (fourth panel). 

A similar behaviour is observed in the mean vertical motions $V_Z$ (Figure~\ref{fig:Z_Vz_maps}, third row), which appear to be related to the vertical distortion in Z (see Figure~\ref{fig:Z_Vz_maps}, second row, and later discussion on the connection between Z and $V_Z$).

A qualitative impression from Figure~\ref{fig:Z_Vz_maps} already gives a taste of the response of the disc to different interaction regimes, but a quantitative analysis is needed to understand the physical mechanisms governing the evolution of the disc under the repeated interactions with a satellite. In the following Sections, we will concentrate on the vertical dimension, quantitatively analyze its temporal evolution and discuss its connection with Sgr's orbit. 

\subsection{Fourier analysis of vertical oscillations}

In this Section, we quantify the vertical response of the disc by means of a Fourier analysis. For a given ring at Galactocentric radius R, we express the mean Z as a sum of infinite Fourier terms terms in Galactic azimuth $\phi$:
    \begin{equation}
        Z(R,\phi) = A_0(R)+ \sum_{m=1}^{\infty} A_m(R) \cos(m \phi - \Phi_m(R))   \quad,
        \label{Fourier_decompos}
    \end{equation}
where $A_0(R)$ is the amplitude of the Fourier term $m=0$ at a given R, which is a constant with respect to $\phi$, while $A_m(R)$ and $\Phi_m(R)$ are, respectively, the amplitude and phase of the vertical terms $m \geq 1$. The azimuthal angle $\phi$ is taken as positive in the direction of Galactic rotation. Rings in R are created in such a way that they contain the same number of particles (40 000), implying that they are more concentrated in the inner regions, and less dense in the disc outskirts. This causes the most external ring to be at $\simeq 23$ kpc, beyond which the requested number of particles cannot be reached.

In this paper, we follow the method used by \cite{2018MNRAS.480.4244C, 2017MNRAS.472.2751C} to study m=1 bending waves in an isolated Milky Way-like disc and under the continuous bombardment of sub-halos. However, our approach is somewhat complementary to their analysis, as we analyse the response of the disc to a single massive Srg-like perturber, with the specific goal of identifying typical behaviours in different dynamical regimes of interaction.   

Figure \ref{fig:Fourier_decomposition} illustrates the power of Fourier decomposition. For a given radius R, every term $m$ can be summarized by either one (for $m=0$) or two (for $m\geq1$) numbers, indicating, respectively, the amplitude, or amplitude and phase. This makes the approach particularly convenient, because we can reduce to some extent the degree of complexity observed in the disc vertical response (first column) by applying a Fourier decomposition, and analyzing the temporal evolution of the amplitudes and phases of the single terms separately (second, third and fourth column). As will become apparent in the following, this approach allows us to identify several features, which otherwise would have been buried within the complexity of the disc response. 

For practical reasons, we first aim to understand whether some Fourier terms are more relevant than others, so that we can devote our attention to the dominant ones, and observe their evolution over time. Figure \ref{fig:amp} shows which Fourier vertical term is dominant at a given time and a given distance from the Galactic center. Vertical dashed lines show Sgr's plane crossings. The plot summarises in a compact manner the evolution of the disc: a vertical slice corresponds to a single snapshot of the simulation, while a horizontal slice shows the temporal evolution at a given annulus at R in the Galactic disc. We can clearly see that the most relevant terms are $m=0,1$ and 2. In contrast, the mode $m=3$ makes a minor contribution, and higher terms are completely absent. Based on the above considerations, in the rest of the paper we will focus on the first three terms, $m=0,1$ and 2.

\subsubsection{Overview of the responses in different regimes \label{Sec:Amp}}

The separate evolution of the $m=0,1$ and 2 amplitudes is shown in Figure \ref{fig:amp_sep}. As we can see, in general, interactions with Sgr are typically associated with a significant increase in the amplitude, for all three terms. Then, the amplitude oscillates a few times, before going back to a semi-quiescent phase. The oscillatory behaviour induced by the interaction with Sgr is more significant and lasts longer in the outer parts of the disc than in the inner regions. 

In the last part of the simulation, the dynamical time-scales of the satellite's orbit are not long enough for the outer disc to recover from each interaction. This effects manifests itself as a rapid increase after one interaction, followed by the above-mentioned oscillatory behaviour, but with a typically very short (Regime 2) or event absent (Regime 3) quiescent phase before another impact occurs.

It is also worth noting that not all plane-crossings have the same impact on the disc's behaviour. For example, during the plane-crossings at $\approx 4.9$ and $6.1$ Gyr, Sgr is far from pericenter (Figure~\ref{fig:regimes}), and the disc reaction is less relevant (or even irrelevant) compared to the other cases.

Finally, we note that the repeated impacts of Sgr on the Galactic disc cause a general amplitude increase of the Fourier vertical terms m=0,1 and 2 throughout the simulation, implying typically stronger vertical distortions in the more perturbed regimes (Regime 2 and 3) than in the quiescent phases.

\subsubsection{Fourier vertical term m=0}

Unlike the other Fourier terms (where $A_m(R) > 0$ by definition), the amplitude of $m=0$ can be either positive or negative. Figure~\ref{fig:m0_amp_sign} repeats the top panel of Figure \ref{fig:amp_sep}, but this time color-coded by sign. In the outer parts of the disc, oscillations are clearly visible. Late regimes (Regime 2 and 3) exhibit amplitudes up to 2 order of magnitudes larger than the early ones (Regime 0 and 1). 

In response to an interaction, the $m=0$ term manifests itself as a series of ring-like vertical perturbation of opposite sign, typically propagating from the inner to the outer disc, similar to a stone causing ripples to propagate out in all directions across the surface of a pond. Ring-like vertical distortions are propagating faster close to an interaction, and become much slower or almost static in quiescent phases. If the sign of $A_0$ is constant for all radii and the amplitude increases with R (e.g. vertical monochromatic features at $\approx$ 4.6 Gyr or 5.7 Gyr), the $m=0$ mode forms an U-shaped bending mode \citep[e.g.][]{1998A&A...337....9R}. This can happen in quiescent phases or in the immediate proximity of an interaction (before evolving into the above-mentioned outgoing vertical perturbations). 

A visual representation of the $m=0$ temporal evolution can be found in Figure~\ref{fig:Fourier_decomposition}, \ref{fig:Fourier_decomposition_appendix1} and \ref{fig:Fourier_decomposition_appendix2}.

\begin{figure*}
	\includegraphics[width=16cm]{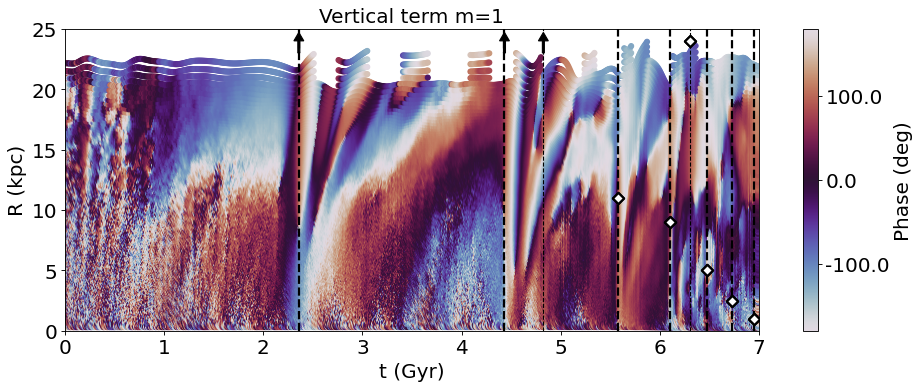}
    \caption{Evolution of the phase angle for the m=1 Fourier term. Dashed vertical lines, arrows and white diamonds are as in Figure \ref{fig:amp}. A zoom of the last region is given in the Appendix (Figure \ref{fig:zoom_amp}). 
    }
    \label{fig:phase}
\end{figure*}

\subsubsection{Fourier vertical term m=1 \label{Sec:phasem1}}

The m=1 term corresponds to a classical S-shape warp \cite[e.g.][]{2019NatAs...3..320C, 2019Sci...365..478S} and might be the dominant Fourier term in the Milky Way \citep{Levine06}. 
Following the approach of \cite{2018MNRAS.480.4244C}, Figure~\ref{fig:phase} shows the the temporal evolution of the m=1 phase angle. Such a plot can be used to determine the morphology of m=1 bending waves in the disc. Provided that the Galactic azimuth $\phi>0$ is positive in the direction of Galactic rotation, one can dissect the plot horizontally, and observe the variation of the phase angle for a fixed radius R as a function of time. Hence, one can deduce if the phase angle (i.e. line-of-nodes of the warp) is moving in a prograde (towards $\phi > 0$) or retrograde (towards $\phi < 0$) direction with respect to Galactic rotation. Alternatively, one can dissect the plot vertically, and determine whether the line-of-nodes of the warp is straight (i.e. constant with R), leading (i.e. $\phi$ increases with increasing radius R) or trailing (i.e. $\phi$ decreases with increasing radius R).


 After an initial relaxation time due to the non-equilibrium initial conditions, a clear coherent pattern manifests itself between $t \approx 0.9$ and $\approx 2.1$ Gyr. At $R \approx 15$ kpc, the phase angle progressively changes from $\approx -45$ deg (dark blue color) to $-170$ deg (light blue). A similar behaviour is observed at all Galactic radii, but especially in the outer parts, with the phase angle slowly moving towards $\phi < 0$, thus opposite to Galactic rotation. Moreover, for a given time, we can take a vertical slice of the plot, and observe that the phase angle varies towards $\phi > 0$ for increasing Galactic radii R, forming a leading spiral. For example, at $t \approx 2$ Gyr the phase angle is $ \approx -180, -140$ and $-100$ deg, respectively, at $ \approx  12, 16$ and 20 kpc. It is worth noting that this pattern is characterized by a long time-scale, which causes the line-of-nodes to vary slowly, with a period much larger than 1 Gyr. This behaviour has been extensively discussed in the literature \citep{2006MNRAS.370....2S, 2016MNRAS.456.2779G, 2018MNRAS.480.4244C}. 


The first plane-crossing of Sgr brings the above-described pattern to an end. The interaction with Sgr resets the phase angle at all radii, and causes the outer parts of the disc ($R \gtrapprox 8$ kpc) to precess (term used in literature to indicate the variation of the line of nodes) in a fast and almost rigid way in the same (prograde) direction with respect to Galactic rotation. At $15$ kpc, the prograde behaviour is maintained for about $200$ Myr. Right after, abrupt variations of the phase angle take place, corresponding to the oscillatory behaviour observed in Section~\ref{Sec:Amp} (i.e. disc flapping). Visually, $m=1$ prograde distortions with a different orientation (i.e. phase angle) with respect to those in the outer parts appear to originate from the inner parts and move to the outer parts of the disc. In contrast, the inner parts of the disc ($R \lessapprox$ 8 kpc) exhibit a retrograde precession, which is progressively restored also in the outer parts of the disc, as soon as the perturbation caused by the interaction is damped over the course of time. Therefore, Regime 1 is characterized by a slow recovery from the first interaction with Sgr, where the disc progressively goes back to a slow, retrograde, differential pattern speed. Also in this case, the line-of-nodes progressively winds up into a leading spiral. 



The second plane-crossing of Sgr (which, in this case, approximately coincides with its second pericentric passage) marks the end of Regime 1. From this point on, time scales become faster, and features are often not coherent over all Galactic radii. It is therefore convenient to zoom in the last region, as shown in Figure \ref{fig:zoom_phase} (top). 

At about 4.5 Gyr, after Sgr' second plane-crossing, most of the disc ($R > 5$ kpc) exhibit a fast, prograde and almost rigid precession, in agreement with what was observed during the first interaction. Between 10 and 20 kpc, the average period (i.e. time required for the line-of-nodes to make one complete rotation around the Galactic center) is of about 300 to 400 Myr, corresponding to an angular speed of about $20$ km/s/kpc. (This information can be approximately obtained from Figure~\ref{fig:zoom_phase} by estimating the time difference between two recurrences of the same phase angle. However, pattern speeds will be more extensively discussed in the following Sections.) 


Between 5.5 and 6.1 Gyr, the outer parts of the disc respond to the impact of Sgr in a similar way to the interactions described before, while the inner parts seem to develop the slow, retrograde pattern observed in Regimes 0 and 1 after about 5.7 Gyr. This situation represents an illustrative example of how different regimes may co-exists at the same time, due to different dynamical time-scales in the Galactic disc. Indeed, here the time interval between two impacts is of the same order of the disc dynamical times in the outer disc (Regime 2), but still larger than those in the inner disc (Regime 1).

After 6.1 Gyr, plane-crossings occur so frequently that the disc is continuosly subjected to external perturbations. As a consequence, complicated and discontinous patterns arise in the disc, being clearly visible also in the inner regions, in which Sgr is progressively being absorbed. In the outer disc ($\approx$ 10 - 20 kpc), a coherent structure survives from about 6.2 to 6.7 Gyr, showing a relatively fast, prograde and nearly rigid precession, completing half of the disc ($180$ deg) in about 300 Myr (roughly corresponding to an angular speed of $\approx$ 10 km/s/kpc). Similarly, other features may survive for a few hundreds of Myr, but they do not remain coherent for more than a few kpc in Galactic radius. Therefore, the recurrent impacts on a short time-scale, together with the proximity of the satellite,
make Regime 3 the most perturbed regime analyzed in this work, dominated by complex disc patterns, which do not remain coherent for very long, both in space and time.

A visual representation of the $m=1$ slow retrograde pattern, typical of quiescent phases, can be seen in Figure~\ref{fig:Fourier_decomposition_appendix1}. The $m=1$ prograde pattern, which occurs during an interaction, can be visualised in Figure~\ref{fig:Fourier_decomposition} and \ref{fig:Fourier_decomposition_appendix2}.





\begin{figure*}
	\includegraphics[width=16cm]{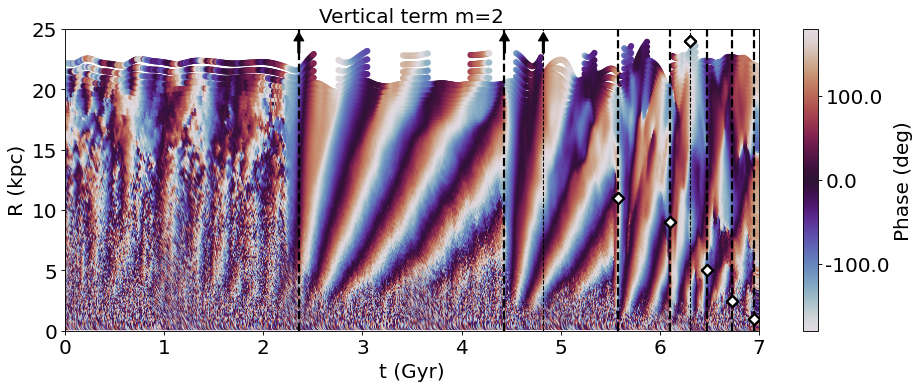}
    \caption{Evolution of the phase angle for the m=2 Fourier term. Dashed vertical lines, arrows and white diamonds are as in Figure \ref{fig:amp}. A zoom of the last region is given in the Appendix (Figure \ref{fig:zoom_amp}). 
    }
    \label{fig:phase2}
\end{figure*}

\begin{figure*}
    \includegraphics[width=16cm]{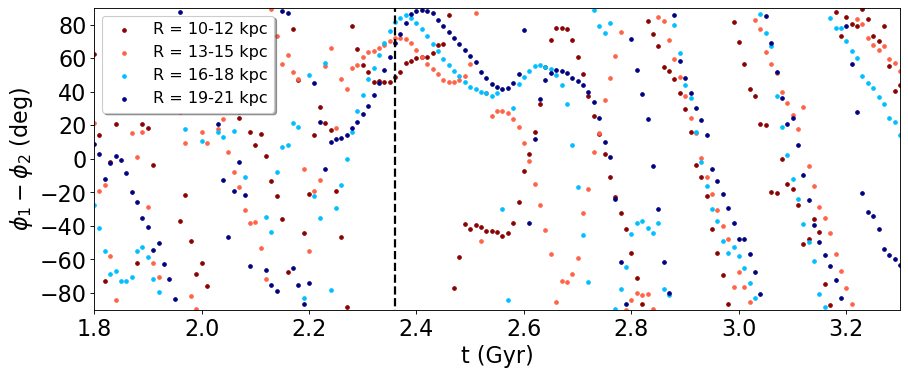}
    \includegraphics[width=16cm]{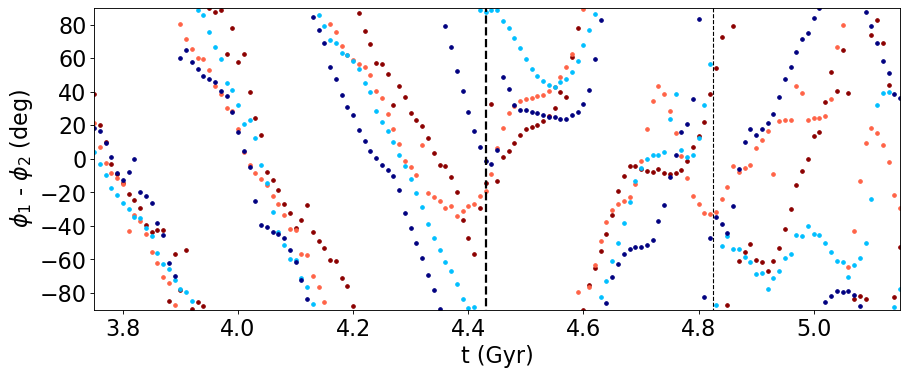}
    \includegraphics[width=16cm]{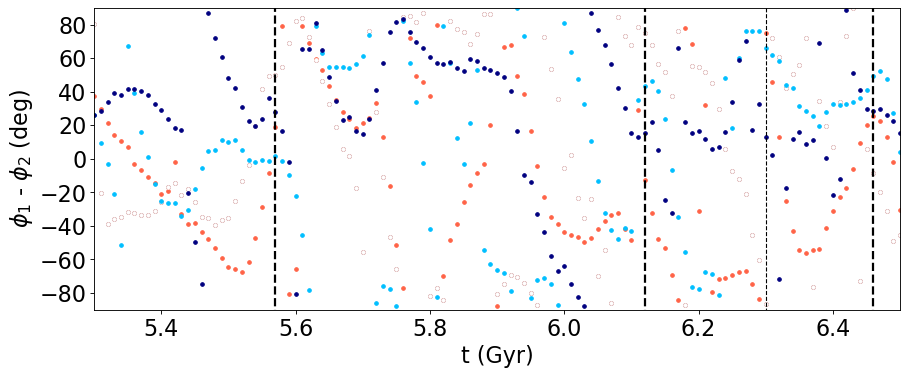}
    \caption{Difference between the phase angle of the m=1 and m=2 Fourier terms at different Galactocentric radii R. Dashed vertical lines as in Figure \ref{fig:amp}. Close to interactions, the two terms are moving at the same speed (i.e. where the difference is constant), illustrating the coupling of the Fourier terms. In the bottom panel, the phase angle inside 12 kpc are represented as open symbols, to avoid crowding.
    }
    \label{fig:phase_diff}
\end{figure*}

\subsubsection{Fourier vertical term m=2}  \label{Sec:phasem2}

The m=2 term corresponds to a bi-symmetric vertical distortion of the Galactic disc. In general, its temporal evolution presents some relevant differences from the previously discussed $m=1$. For example, $m=2$ is completely absent in the pre-interaction regime, and gets excited by the first plane crossing of Sgr (at $\approx$ 2.3 Gyr), as shown by Figure~\ref{fig:amp_sep}. 


The temporal evolution of the $m=2$ phase angle is shown in Figure \ref{fig:phase2} (see also Figure \ref{fig:zoom_phase}, which is zoomed in the most perturbed region). As we can see, the $m=2$ term exhibits a very regular behaviour. Even in the latest regimes, where Sgr's impacts become more frequent and the disc is strongly perturbed, the $m=2$ pattern can be roughly recognised. For comparison, the m=1 term appears to be much more fragile and easily disrupted. 

As will be better explained in the spectral analysis, the $m=2$ bending waves typically move in a prograde direction (also in quiescent phases, unlike $m=1$). Their angular speeds show a very mild variation with radius, which causes the bending waves to wrap up very slowly. Starting from an initially long radial wavelength ($>25$ kpc), the phase of the m=2 term progressively winds up, reaching a radial wavelength of $\approx$ 8 kpc almost 2 Gyr after the latest interaction. Therefore, the pitch angle of the m=2 vertical wave can, in principle, be used as a clock to date the most recent interaction.

During a plane-crossing, if that also coincides with a pericentric passage, the phase-angle of the m=2 term gets reset at all radii, and, intially straight, moves in a prograde direction, similarly to what observed for the $m=1$ term. Figure \ref{fig:phase_diff} shows the difference between the phase angle of the $ m=1$ and the $m=2$ term ($\phi_1$ and $\phi_2$, respectively) at different times. 
During the first two pericenters, the $m=1$ and the $m=2$ term seem to be somewhat aligned, moving at approximately the same speed (i.e. their difference is roughly constant). Such behaviour might be due to a coupling between the Fourier terms, which occurs as a response to the interaction with Sgr. However, this effect becomes less clear in the latest and more perturbed regimes, where the presence of Sgr inside the disc does not allow coherent features to survive for more than a few kpc in Galactic radius. In contrast, very regular patterns can be observed in quiescent regions (e.g. $ 3 < $ t $  < 4.4$ Gyr), where the two Fourier terms revolve at their own typical speed, with the $m=1$ being slightly retrograde (see above) and the $m=2$ typically prograde. This difference is causing the difference $\phi_1 - \phi_2$ to be decreasing with time. 


\begin{figure*}
	\includegraphics[width=15cm]{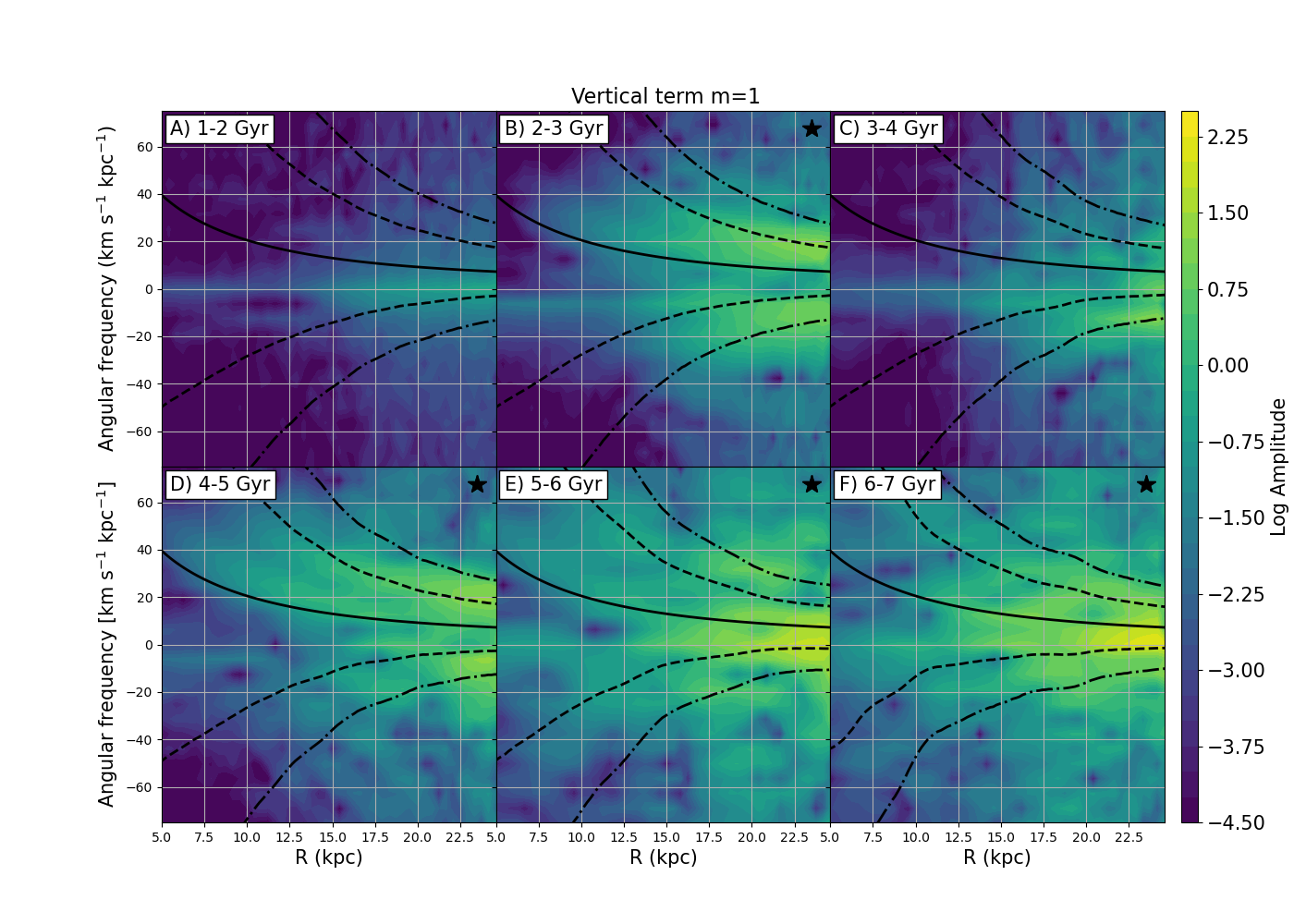}
    \caption{Vertical spectrograms of the Fourier m=1 vertical term. 
    Each panel covers a temporal bin of 1 Gyr in the simulation. Panels marked by a black star are those where one (or more) interaction(s) with Sgr took place. The solid curve shows corotation. The dashed curves show the vertical resonances at $\Omega(R) \pm \nu_{z}(R)$, where $\Omega(R)$ is Galactic rotation and $\nu_{z}(R)$ is the vertical frequency (see Appendix). The dashed-dotted curves show the vertical resonances at $\Omega(R) \pm 2 \nu_{z}(R)$,
    }
    \label{fig:vertspect_m1}
\end{figure*}

\begin{figure*}
	\includegraphics[width=15cm]{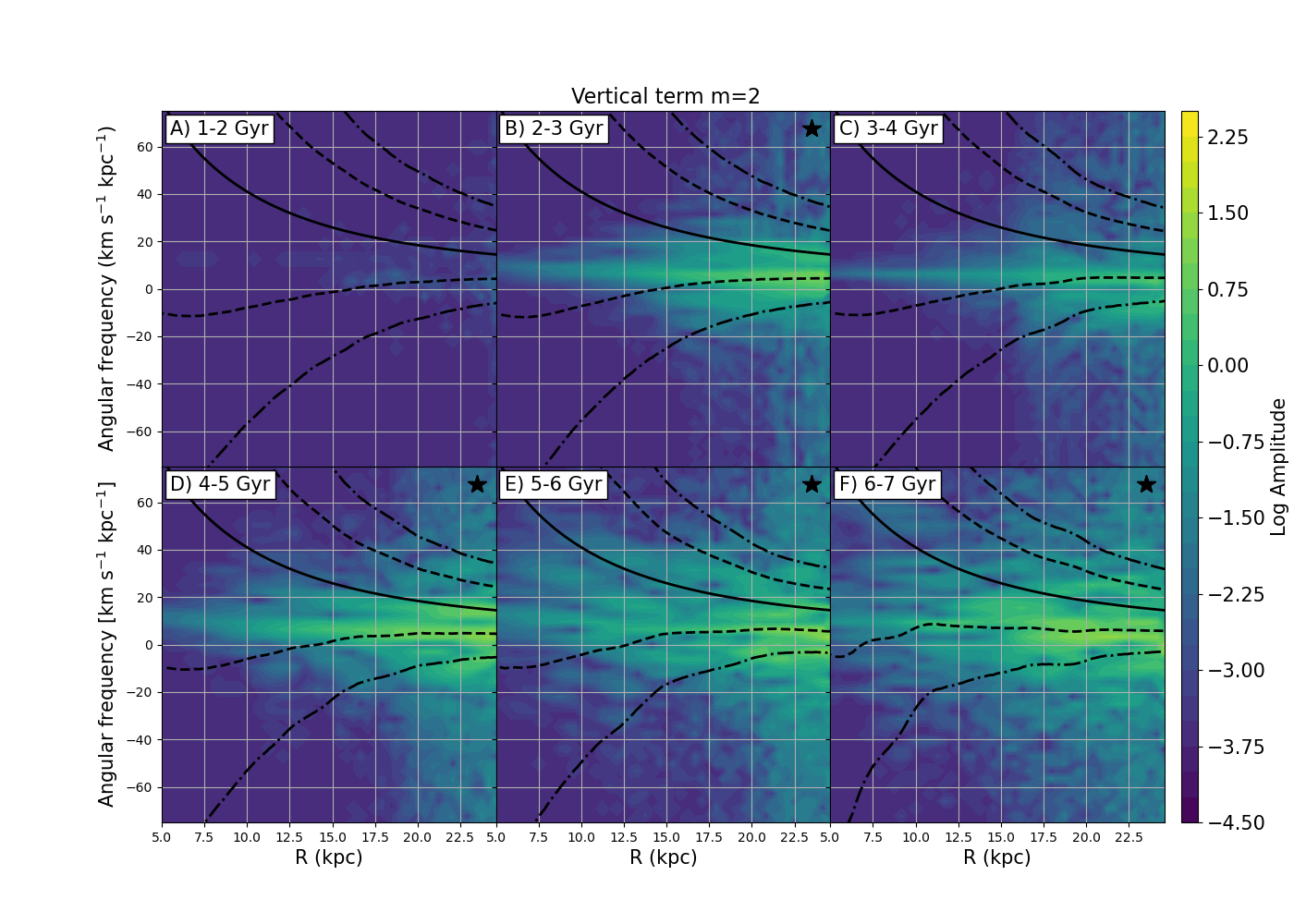}
    \caption{Same as Figure \ref{fig:vertspect_m1}, but for the Fourier m=2 vertical term. 
    }
    \label{fig:vertspect_m2}
\end{figure*}

\subsubsection{Spectral analysis} \label{sec:spec_analysis}
Originally developed for density waves  \citep{1986MNRAS.221..195S, 1997A&A...322..442M,2011MNRAS.417..762Q,2020ApJ...890..117D}, spectral analysis has been recently extended to bending waves \citep{2018MNRAS.480.4244C, 2017MNRAS.472.2751C}, allowing us to describe them in terms of frequency power spectrum as a function of Galactocentric radius. A full description of vertical spectrograms can be found in \citet{2018MNRAS.480.4244C, 2017MNRAS.472.2751C}, and technical details are given in Appendix \ref{sec:appendix_vert_spec}. In this paper, we will use vertical spectrograms to identify the most relevant features over a given time baseline. Specifically, we will be interested in determining the angular speed (i.e. temporal variation of the phase angle $\Phi_m(R)$, see Equation \ref{Fourier_decompos}) of the most relevant Fourier terms $m \geq 1$ in different regimes of interaction with Sgr. 

Figure \ref{fig:vertspect_m1} consists of six vertical spectrograms, showing the frequency spectrum for $m=1$ bending waves in different periods of the simulation. To relate the vertical response of the disc to the satellite's orbit, we chose a temporal baseline of 1 Gyr for each spectrogram, which implies a resolution on the vertical axis of 6.28 km/s/kpc (the temporal resolution of the simulation is 10 Myrs). (For a sharper resolution on the vertical axis, implying a longer baseline, see Appendix \ref{sec:appendix_vert_spec}.) One of the most obvious features of Figure \ref{fig:vertspect_m1} is that the power reaches typically larger values (i.e. stronger bending waves) in the spectrograms where one (or more) interaction(s) with Sgr occurred (panels marked by a black star), whereas it remains relatively low in quiescent regions (panels A and C).

It is also instructive to look at where the power is concentrated at different stages of the Milky Way-satellite interaction. In quiescent regions, most of the power is concentrated close to angular speed 0 (panel A), or slightly retrograde values (panel C), in the outer regions of the disc. This approximately aligns with the vertical resonance $\Omega - \nu_z$, where $\Omega$ is the angular rotation rate of the Galaxy and $\nu_z$ is the vertical epyciclic frequency. This is the so-called slow wave, and is expected to be retrograde because, in a flattened galaxy, the vertical frequency $\nu_z>\Omega$. The slow and retrograde bending wave has been already discussed in the literature \citep[][]{2008gady.book.....B, 2017MNRAS.472.2751C, 2018MNRAS.480.4244C}.

However, in panel B, power is located in two main regions: one close to $ \simeq$20-25 km/s/kpc, indicating the prograde precession associated with Sgr's impact, and the slow retrograde precession, which is visible before/after the effects of the perturbation have relaxed. The prograde feature disappears in panel C, where no interaction take place. Therefore, there is a clear connection between the vertical pattern speeds in the Galactic disc and the satellite's orbit. This is further confirmed by panels D, E and F, where multiple interactions are present, and transient prograde features become prominent. We also observe that prograde angular speeds in panel E and F are typically slower than those in panel B and D. 

Figure \ref{fig:vertspect_m2} repeats the vertical spectrograms of Figure \ref{fig:vertspect_m1}, but now for the $m=2$ term. Considerations similar to the ones above can be done here, with the difference that: (i) the m=2 term is completely absent before the first interaction (panel A), and (ii) the resonance curves are different, as the angular speed $\omega = m \Omega \pm \nu_z$, with $\omega = m \Omega_p$, where $\Omega_p$ is the pattern speed of the bending wave. In the previous sections, we found that $m=2$ bending waves present a very regular behaviour, characterized by a prograde angular speed. Indeed, the spectrograms indicate that, in the outer regions, the power aligns with the vertical resonance at $\omega = m \Omega - \nu_z$ \citep[see ][]{Bland-Hawthorn2020}, which demarcates a prograde angular speed.

\begin{figure*}
    \includegraphics[width=16cm]{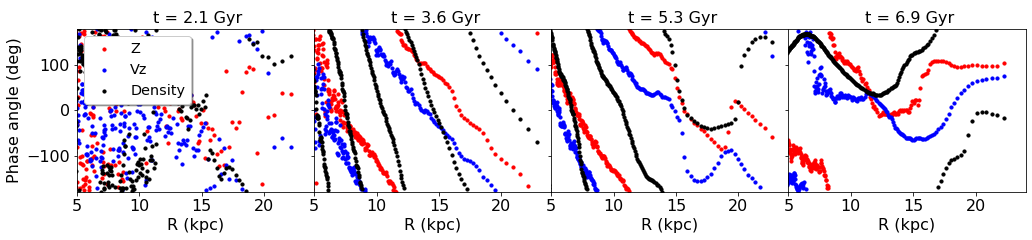}
    \caption{Comparison between the phase angle of the m=2 mode for Z (red), Vz (blue) and density (black) at t=3.3 Gyr, 5.5 Gyr and 6.5 Gyr (from left to right).}
    \label{fig:m2_phase_comparison}
\end{figure*}

\begin{figure*}
	\includegraphics[width=16cm]{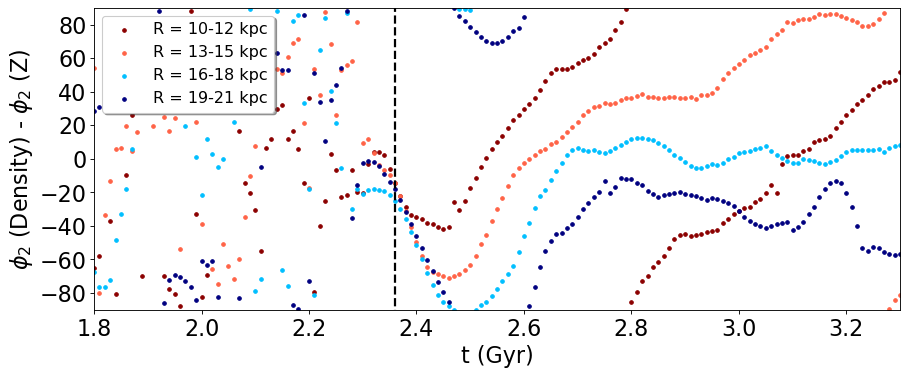}
	\includegraphics[width=16cm]{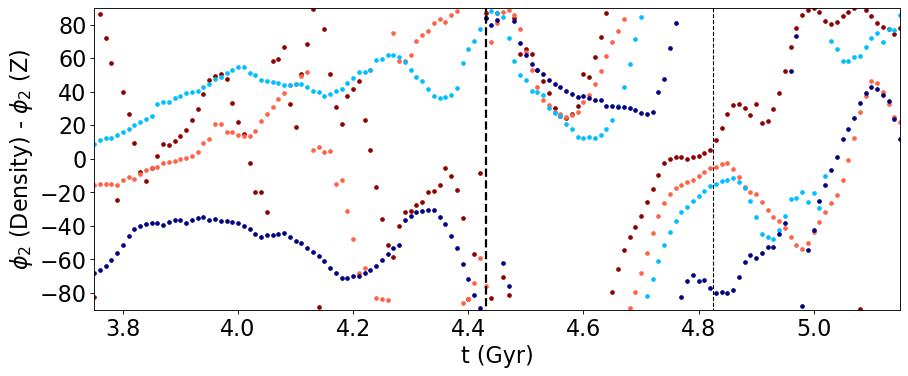}
	\includegraphics[width=16cm]{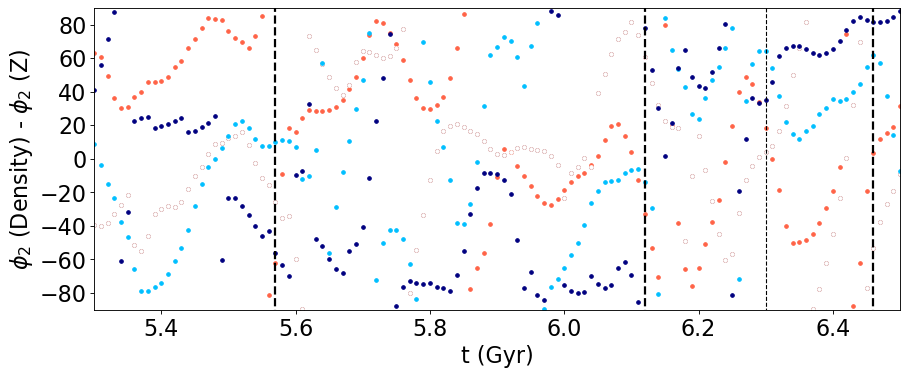}
    \caption{Same as Figure \ref{fig:phase_diff}, but showing the difference between the m=2 density and m=2 vertical Fourier terms.
    }
    \label{fig:phase_diff_m2_dens_Z}
\end{figure*}

\subsubsection{Relating response in different dimensions}


Figure \ref{fig:m2_phase_comparison} shows the phase angle of the $m=2$ vertical term (Z) as a function of Galactocentric radius R, compared to other quantities that might be of interest. The $m=2$ term in vertical velocity ($V_z$) appear to be typically shifted by $90 \deg$ with respect to the $m=2$ phase in Z, as expected from a wave-like motion. 

When comparing the $m=2$ vertical term (Z) and the $m=2$ spiral arms in density, we find that, in general, there is no obvious correlation. This can be also seen in Figure~\ref{fig:Z_Vz_maps}, where the spiral arms in density appear more tightly wound than the ones in the vertical spiral wave. Indeed, such behaviour is somewhat expected, as the $m=2$ density and bending waves have, respectively, a pattern speed of $\Omega(R) - \kappa/2 $ and $\Omega(R) - \nu_z/2 $. This is in good agreement with the recent analysis presented by \cite{Bland-Hawthorn2020}. 

Figure \ref{fig:phase_diff_m2_dens_Z} shows the angular difference between the $m=2$ phase angle in density and in Z as a function of time for different Galactocentric radii. As we can see, the $m=2$ density and bending waves usually move at their own typical rate, with the $m=2$ in density wave being typically faster than $m=2$ in $Z$. This is apparent when the difference $\phi_2$ (Density) - $\phi_2$ (Z) increases as a function of time. However, in proximity of an interaction, the behaviour is different. The $m=2$ density and bending waves tend to align, and move at a similar speed for a very short time. This behaviour might highlight a resonance, caused by the interaction with Sgr. 


\begin{figure*}
	\includegraphics[width=7.8cm]{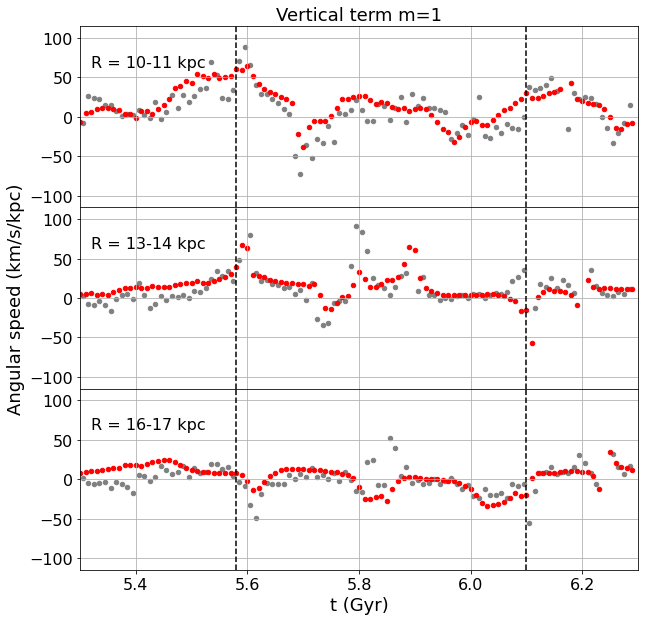}
	\includegraphics[width=9.3cm]{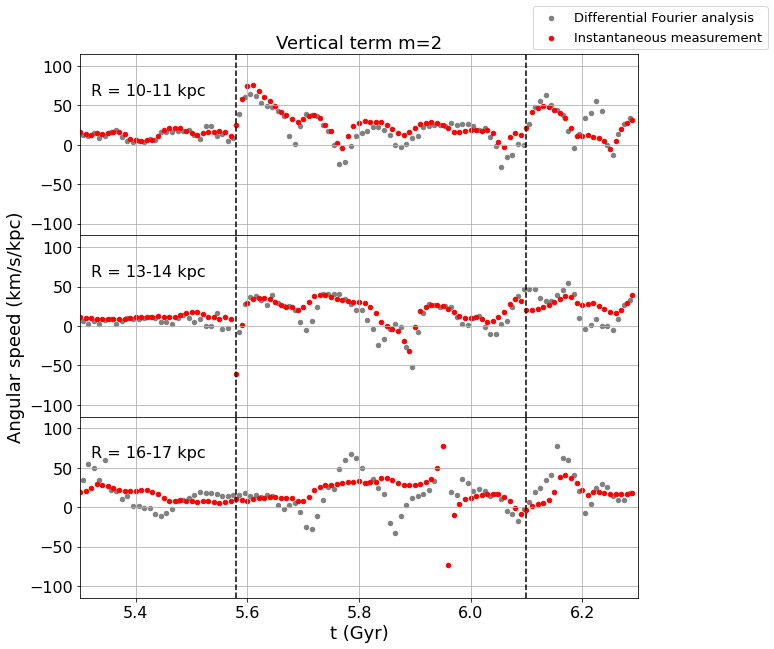}
    \caption{Comparison between the angular speed obtained from a data-like approach (red points, see text) and the actual behaviour of the disc (grey points and shaded area) for the m=1 (left) and m=2 (right) Fourier vertical terms. Different rows refer to different positions in the Galactic disc. The grey points are obtained by simply subtracting the phase angles at two subsequent snapshots (for a more sophisticated approach, see Section \ref{Sec:VertResp}).}
    \label{fig:sims_vs_data}
\end{figure*}

\begin{figure*}
	\includegraphics[width=8cm]{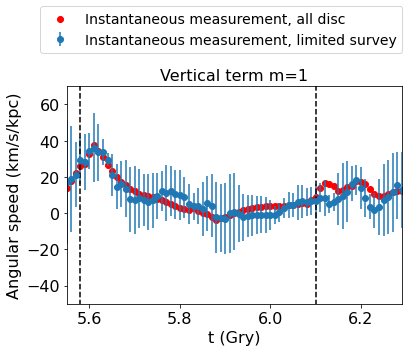}
	\includegraphics[width=8cm]{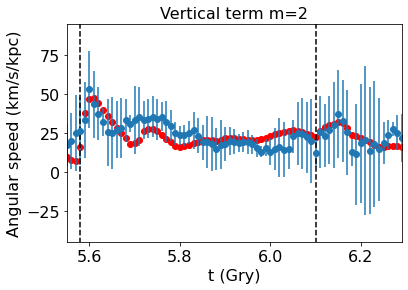}
    \caption{Same as Figure \ref{fig:sims_vs_data}, but comparing the results of the data-like approach using the entire disc (red points) and a limited volume of the Galactic disc (blue points). For a given snapshot, the blue points show the median angular speed obtained from eight different volumes (see text), while the error bar shows the median absolute deviation (MAD) of the obtained values.  }
    \label{fig:limited_survey}
\end{figure*}


\section{Observing the disc response } \label{Sec:obs_disc_response}
The previous Section showed that the evolution of the Galactic disc and the orbit of a perturbing satellite are tightly correlated, thereby leading to various expectations for vertical pattern speeds in the Galactic disc and other quantifiable properties, depending on the status of interaction. In principle, the response of the disc can be inferred from observations, and used to learn more on the interaction with the perturbing satellite(s) and the perturbed halo, with the caveat that other internal mechanisms should be considered as well, such as the bar buckling (with the further caveat that we do not have an indication of when in the lifetime of the Galaxy this may have occurred). 

The evolution of the disc in N-body simulations can be deduced by comparing subsequent snapshots at different times (as was done in the previous Section). However, in real data, observations provide us with positions and velocities of stars in the Galaxy only at the present time. Therefore, a method is required to infer the actual behaviour of the disc using only the instantaneous measurements of the 6-dimensional phase-space of stars.


In this Section, we aim to mimic the real data approach, in order to understand whether it is possible to recover the actual instantaneous behaviour of the Galactic disc by simply analysing positions and velocities of stars at a given time. Certainly, the equations governing the vertical response of the Galaxy to the recurrent impacts of a satellite galaxy should be based on a deep dynamical investigation, which is beyond the scopes of the present paper. Instead, here we use a very simple kinematical approach, and test its validity and limitations. The approach currently includes the three Fourier terms $m=0,1$ and 2 (Appendix \ref{sec:appendix_visualize_fourier} shows that this is a reasonable approximation), but it can be easily extended to other Fourier terms, if needed. We model the disc using an exponential vertical profile, whose vertical displacement is determined as follows. For a given R, the amplitude $h_0$ of the $m=0$ term is free to change with time. The two terms $m=1$ and $m=2$ have a time-varying amplitude ($h_1$ and $h_2$, respectively) and phase ($\phi_1$ and $\phi_2$, respectively).
The above time-dependent vertical displacement has implications on the vertical velocities of the stars, whose mean trend can be calculated using the Collisionless Boltzmann Equation \cite[see Equation 6 and its derivation for the $m=1$ term in][]{2020NatAs...4..590P}:
\begin{equation}
\begin{split}
    V_z(R,\phi)= \frac{\partial h_0}{\partial t} + \Bigl( \frac{V_{\phi}}{R} - \frac{\partial \phi_1}{\partial t} \Bigr) h_1 \cos(\phi - \phi_1) + \frac{ \partial h_1}{ \partial t} \sin(\phi - \phi_1) \\
    + \Bigl( 2 \frac{V_{\phi}}{R} - \frac{\partial \phi_2}
    {\partial t} \Bigr) h_2 \cos(2\phi - \phi_2) + \frac{ \partial h_2}{ \partial t} \sin(2\phi - \phi_2)
	\label{eq:Vzmode2} \quad,
\end{split}	
\end{equation}
where the kinematic parameters $\frac{\partial \phi_1}{\partial t} $ and $\frac{\partial \phi_2}{\partial t} $ are the angular frequencies of the $m=1$ and $m=2$ terms, respectively, and $\frac{\partial h_0}{\partial t} $, $\frac{\partial h_1}{\partial t} $, $\frac{\partial h_2}{\partial t} $ are the amplitude variations of the corresponding Fourier terms. These five parameters are, in general, non-zero, and can be statistically inferred using positions and velocities ($R, \phi, V_{\phi}, V_z$) of the stars in the Galactic disc at a given time using Equation~\ref{eq:Vzmode2}. To do this, in this experiment we use a simple non-linear least squares method, assuming that the (spatial) amplitudes and phases of the $m=0,1$ and 2 terms are known. Hereafter, we will refer to this method as the \emph{Instantaneous measurement}, since the kinematic parameters are recovered from the 6-dimensional phase-space information for stars at a fixed time, as in the case of real data.

As anticipated above, an alternative way to determine the instantaneous evolution of the Galactic disc is to compare the amplitudes and phases at different snapshots. For example, one can calculate the angular speed $\frac{\partial \phi_1}{\partial t} $ at given radius R by naively taking the phase difference at two subsequent snapshots. Alternatively, one might use the vertical spectrograms (see Section \ref{sec:spec_analysis}), which, however, provide us with the power distribution on a relatively large time-scale (1 Gyr), due to the simulation temporal resolution. Therefore, in this simple experiment, we use the first approach, and name it \emph{Differential Fourier analysis}. 

Figure \ref{fig:sims_vs_data} shows the results obtained when recovering the angular speeds $\frac{\partial \phi_1}{\partial t} $ and $\frac{\partial \phi_2}{\partial t} $ of the $m=1$ and $m=2$ terms according to the Instantaneous measurement and the Differential Fourier analysis at different Galactic radii. As we can see, notwithstanding its simplicity, our data-like approach seems to produce results in rough agreement with the Differential Fourier analysis, even in the latest regimes, where the disc is more perturbed.


However, a realistic magnitude limited survey does not cover the entire Galactic disc. Therefore, we perform an additional experiment, to test the applicability of the data-like approach on a limited portion of the Galactic disc. We place the simulated Sun at 12 kpc from the Galactic center, to approximately reproduce a realistic warp amplitude based on the observed mean vertical displacement at a given heliocentric distance \citep[e.g.][]{2006A&A...451..515M}. While the Sun's distance from the Galactic center is maintained fixed, we place the Sun at 8 different azimuthal angles (spaced by 45$^o$) for a given snapshot. Then, we select only stars within 10 kpc from the Sun, to approximately mimic a Gaia DR3-like sample of stars. For each snapshot, we end up with 8 mock catalogs of stars, containing $\simeq 600 \, 000$ particles (which implies a lower statistic than the one that can presumably be reached with Gaia DR3). Due to the low statistics, in this second experiment we fit for a single value of angular speed $\frac{\partial \phi_1}{\partial t} $ and $\frac{\partial \phi_2}{\partial t} $, instead of evaluating each radial ring separately. (This approximation is equivalent to averaging over the portion of the disc covered by the mock dataset.) Finally, we take the median of the results from the 8 mock catalogs, and calculate the dispersion around the median value, and repeat the procedure for all the snapshots of interest. The results are presented in Figure \ref{fig:limited_survey}. As we can see, the dispersion can be significant in some places, as expected, due to the complexity of the physical mechanisms at work in our Galaxy, which are presumably not captured by our simple approach. However, the median value is in good agreement with the values obtained using the entire disc, meaning that the approach is unbiased.



\section{Discussion and conclusion}  \label{Sec:summary}



\subsection{Key results}
Using N-body simulations from \citetalias{Laporte18}, we have described the interaction between the Milky Way and Sgr in terms of four different interaction regimes (called Regime 0,1,2,3), based on the orbital history of Sgr and the dynamical time scales in the Galactic disc. Each regime dictates the properties, extent and evolution of the disc perturbations. We have quantitatively characterised the evolution of the vertical distortions of the disc under the repeated impacts of Sgr using Fourier decomposition and spectral analysis. 
Some of the observed features have been already discussed in literature, while others are new. Finally, we have tested the validity of pattern speed measurements in a realistic model.
The main results of the paper are summarized below:
\begin{enumerate}
    \item  The $m=0$ term manifests itself as a ring-like vertical distortion with alternate sign propagating from the inner to the outer parts of the disc. The radial propagation speed is faster after an interaction, and becomes much slower (almost static) in quiescent phases. 
    \item The $m=1$ term is characterized by a slow, differential, retrograde precession
    in the quiescent regimes, while it moves in a relatively fast and prograde direction in the last regimes or close to an interaction.
    \item The $m=2$ term is excited by the first interaction with Sgr, and presents a very regular behaviour, apparent as a prograde bi-symmetric trailing spiral, which progressively winds up with time after an interaction.
    \item In quiescent phases, the Fourier vertical terms $m =1$ and $m=2$ revolve at their own typical rate, which is determined by vertical Lindblad resonances. However, their angular frequency is forced to change during the interactions with Sgr, and they both move in a prograde direction for a few hundred Myrs, at approximately the same speed. 
    Such coherent motion might perhaps indicate a resonance, which is arising as a response of the disc to the interaction with the satellite. 
    \item The instantaneous evolutionary state of the Galactic disc can be parameterised using the amplitude and phase angle variation (angular speed) of the Fourier terms. Such parameters can be recovered in an unbiased way from stellar positions and velocities in the Galactic disc at a given time (i.e. similar to the real data case). 
\end{enumerate}

\subsection{Implications}

In this work, we have shown that the $m=0, 1$ and 2 Fourier vertical terms dominate and behave in a different way in distinct phases of the interaction process with a satellite galaxy. Although some caveats (discussed below) should be kept in mind, our results imply that \emph{the kinematic parameters describing the instantaneous vertical evolution of the Galactic disc have a constraining power on the dynamical state of the interaction}, when a Milky Way-Sgr like accretion is considered. Taking advantage of current and future datasets, which allow us to characterize the Galactic disc on an unprecedentedly large volume, one can (i) place observational constraints on the evolution of the perturbed Galactic disc, and (ii) use high-resolution N-body simulations to bridge the gap between observation and theory, to suggest a possible dynamical interpretation of the observations in the context of a Milky Way-Sgr interaction.

The scenario analysed in this work provides a possible framework for the interpretation of the measured warp ($m=1$) pattern speed \citep[$\simeq 11 \pm 3$   km/s/kpc, 13.6 km/s/kpc, respectively][]{2020NatAs...4..590P,Cheng2020}, which would be more consistent with the Milky Way currently being in the second or third dynamical regime (see Figure \ref{fig:vertspect_m1}E, \ref{fig:vertspect_m1}F, and \ref{fig:sims_vs_data}) or, in any case, close to an interaction, 
rather than in a semi-quiescent phase (e.g. Figure \ref{fig:vertspect_m1}A and \ref{fig:vertspect_m1}c, where no interactions have occurred in a time-interval of $\simeq 1.5$ Gyr). 

Additional measurements can potentially give further insights into the origin of vertical disturbances and relate to satellite contribution. For example, the pattern speed of the $m=2$ vertical term might be measured using future datasets. Such result would be not only interesting by itself, but also because the difference between the angular speed of the m=1 and m=2 terms represents a complementary way of determining if an interaction has occurred in the last few hundreds of Myrs. Additionally, given that the $m=2$ mode is persistent and winds-up very slowly (see Fig. \ref{fig:phase2}), its pitch angle, even measured locally, may provide a good clock for measuring the time since the last perturbation.

 \subsection{Limitations and future prospects}

This work aims to make further progress towards the understanding of the vertical response of the disc to the recent accretion history of the Galaxy. As a starting point, we analysed the response to the repeated impacts of a Sgr-like satellite. However, a number of caveats should be kept in mind, when interpreting our results in their wider context. For instance, modelling Sgr orbital history from the virial radius crossing to the present-day is notoriously difficult, which is further exacerbated by the large uncertainties on its parameters (e.g. progenitor mass profile, last pericentric passage, etc.), and those of the Milky Way's own potential. 
Hence, instead of aiming to reproduce the \emph{exact} orbital history of Sgr and the \emph{exact} vertical response of the disc, we have rather concentrated ourselves on the exploration of the physical spatial reaction to (and the relative time-scales of) an interaction between a Milky Way-like galaxy and a Sgr-like satellite. In addition to this, it should be noted that the Magellanic Clouds also play a role in generating vertical pattern speeds in the Galactic disc \citep[e.g.][]{Weinberg06,laporte18a}. Its effect in combination with Sgr was already explored in \cite{Laporte18} showing that the recent interaction with the LMC modulates the response due to Sgr which is particularly noticeable in the outer disc ($\geq16$ kpc). In addition to this, we note that the simulation does not include a gaseous component, which might contribute to the bending wave damping. A deeper exploration is left for future work. Finally, it should be mentioned that our work did not consider the effects of the bar buckling; whether it has a significant impact or not on the outer disc should be further explored. Future works comparing observational constrains on the disc vertical evolution with possibly more sophisticated dynamical models will certainly reveal further information on the recent accretion history of the Milky Way.

\section*{Acknowledgements}

EP, CL and ED thank Lawrence Widrow for useful discussions. EP acknowledges financial support from the Center of Computational Astrophysics, Flatiron Institute (New York, USA), which enabled a long-term visit in 2019 to develop this project. CL thanks the Observatoire de la C\^ote d'Azur for hosting him on a long-term visit. This work used the Extreme Science and Engineering Discovery Environment (XSEDE), which is supported by National Science Foundation grant number OCI-1053575. This work was supported in part by World Premier International Research Center Initiative (WPI Initiative), MEXT, Japan. EP acknowledges the partial support of ASI through contract ASI 2018-24-HH.0, Gaia Mission - The Italian Participation to DPAC. CL acknowledges funding from the European Research Council (ERC) under the European Union’s Horizon 2020 research and innovation programme (grant agreement No. 852839). KVJ and DF were supported by NSF grant AST-1715582.

\bibliographystyle{mnras}
\bibliography{example} 
%
%
%
%

\appendix
\section{Visualizing the Fourier decomposition} \label{sec:appendix_visualize_fourier}

\begin{figure*}
	\includegraphics[width=12.5cm]{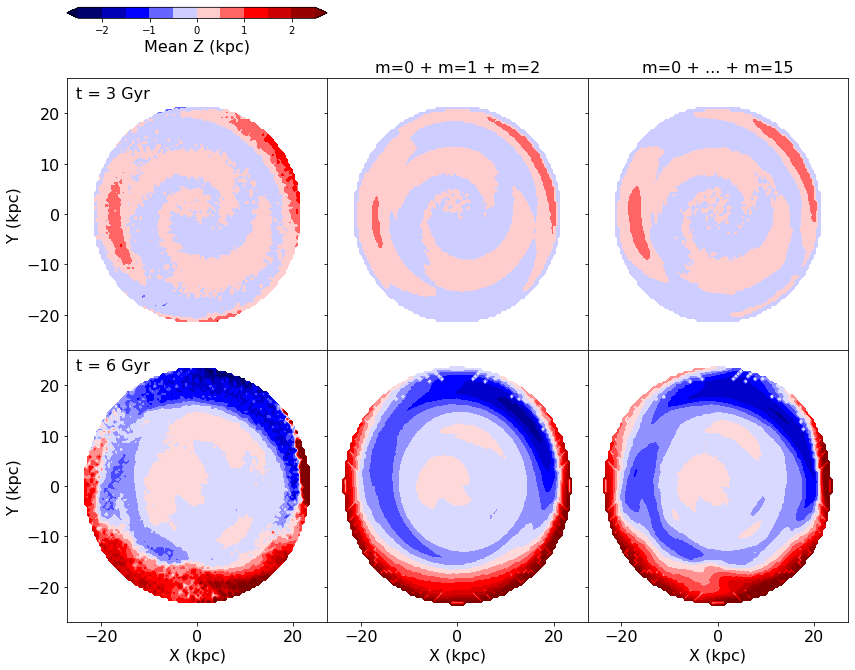}
    \caption{The mean vertical displacement of the Galactic disc (first column) compared to the corresponding Fourier decomposition, using the sum of the first three terms m=0, m=1 and m=2 only (second column) and the first sixteen terms m=0, ..., m=15 (third column). Two snapshots were considered to test the validity of the approximation both in the early stages of the simulation (first column) and the more perturbed regimes (second row).
    \label{fig:Fourier_decomposition_example}}
\end{figure*}

In this Section, we give further details on the Fourier decomposition performed in Section \ref{Sec:VertResp}. As discussed in the text, in this paper we mainly focus on the Fourier terms m=0, m=1, m=2 for practical reasons. Those terms are found to be the most relevant ones (see main text), and their combination is expected to reproduce, to a first approximation, the vertical structure of the Galactic disc. Indeed, Figure \ref{fig:Fourier_decomposition_example} shows that the main features observed in the simulation (first column) are already approximated in an acceptable way by the crude sum of the first three terms (second column), although the approximation is obviously more accurate when higher order terms are included (third column). In this paper, higher order terms are not investigated for practical reasons, and we leave their exploration to future works. 

\begin{figure*}
	\includegraphics[width=12.5cm]{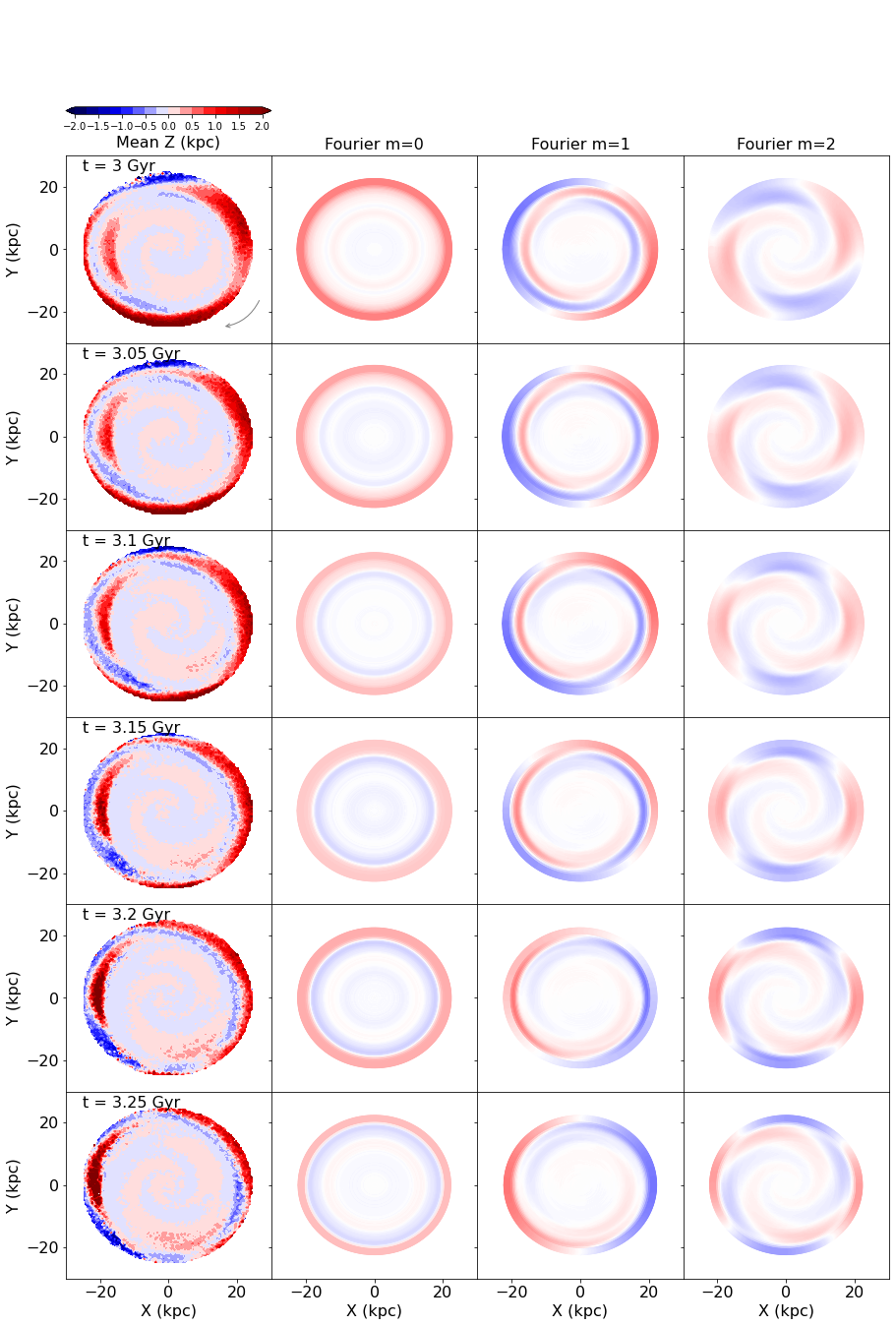}
    \caption{Same as Figure \ref{fig:Fourier_decomposition}, but considering a quiescent phase in Regime 1.
    \label{fig:Fourier_decomposition_appendix1}}
\end{figure*}

\begin{figure*}
	\includegraphics[width=12.5cm]{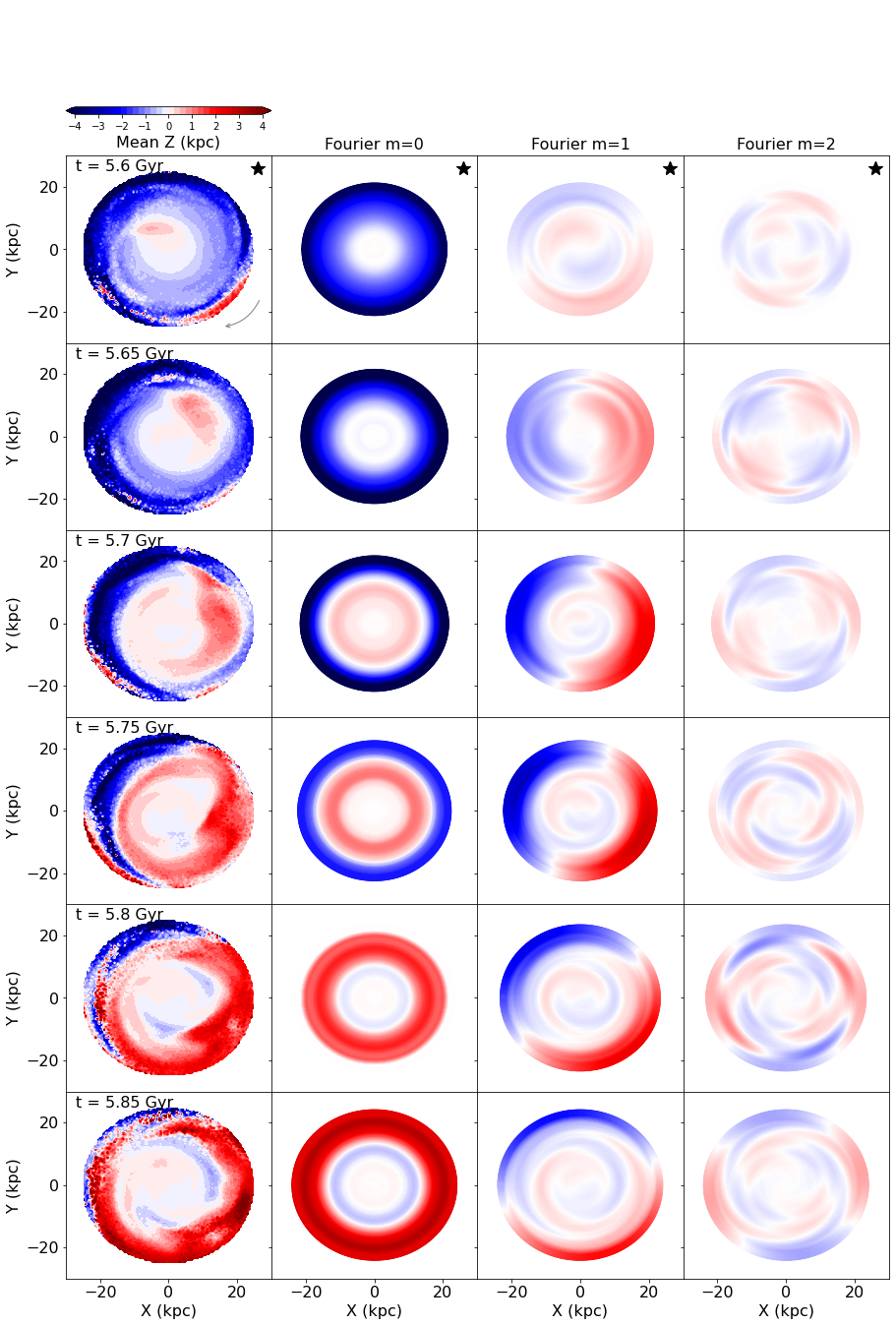}
    \caption{Same as Figure \ref{fig:Fourier_decomposition}, but considering Regime 2.
    \label{fig:Fourier_decomposition_appendix2}}
\end{figure*}

To complement Figure \ref{fig:Fourier_decomposition}, which was dedicated to the first interaction with Sgr, we here present Figure \ref{fig:Fourier_decomposition_appendix1} and \ref{fig:Fourier_decomposition_appendix2}, showing, respectively, the evolution of the three first Fourier terms in a quiescent phase after the first interaction (Regime 1), and after an interaction in the most perturbed regimes (Regime 2). As we can see, m=0 looks almost static in the quiescent phase (Figure \ref{fig:Fourier_decomposition_appendix1}), while manifests itself as fast, outgoing, ring-like vertical distortions close to an interaction (Figure \ref{fig:Fourier_decomposition} and \ref{fig:Fourier_decomposition_appendix2}). The m=1 term moves in a prograde direction close to an interaction (Figure \ref{fig:Fourier_decomposition} and \ref{fig:Fourier_decomposition_appendix2}), but moves in a slow and retrograde fashion in quiescent phases (Figure \ref{fig:Fourier_decomposition}), creating a tightly wound feature. After an interaction, the m=2 term progressively winds up as well, although in Figure \ref{fig:Fourier_decomposition_appendix2}) features are more discontinuos, because of the direct influence of Sgr (which is in the Galactic plane (z=0) at R=11 kpc from the Galactic center at t=5.58 Gyr).

\section{Zoom last region}
\begin{figure*}
	\includegraphics[width=16cm]{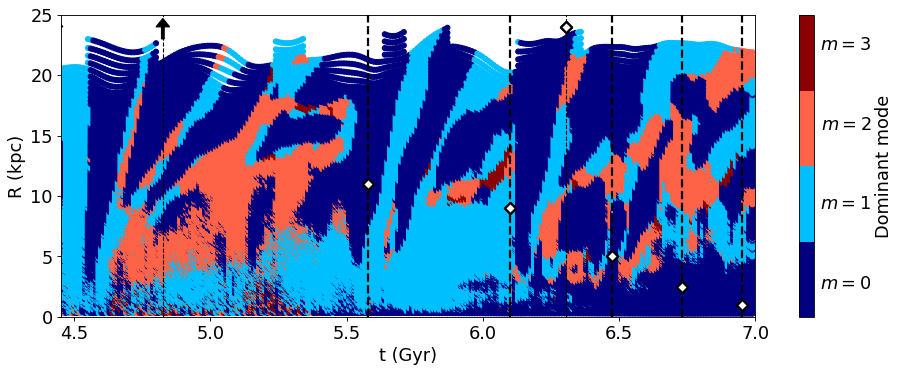}
    \caption{Same as Figure \ref{fig:amp} , but zooming in the most perturbed region.}
    \label{fig:zoom_amp}
\end{figure*}

\begin{figure*}
    \includegraphics[width=16cm]{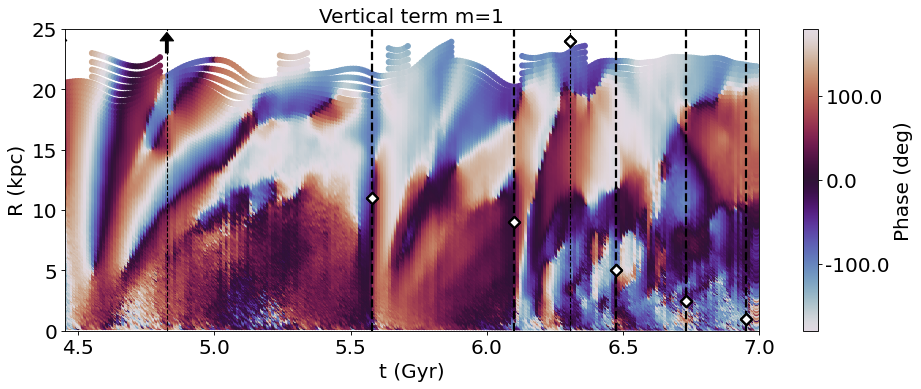}
	\includegraphics[width=16cm]{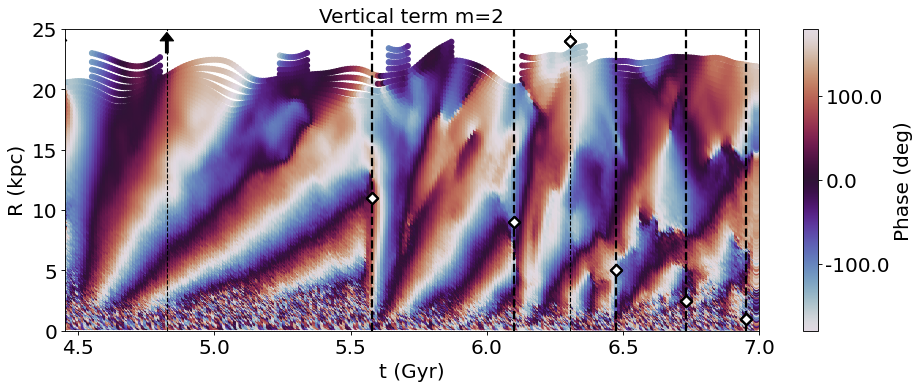}
    \caption{Same as Figure \ref{fig:phase} (top) and Figure \ref{fig:phase2} (bottom), but zooming in the most perturbed region.}
    \label{fig:zoom_phase}
\end{figure*}

In this section, we reproduce Figures \ref{fig:amp}, \ref{fig:phase} and \ref{fig:phase2}, but zooming in the latest and more perturbed regimes (Figure \ref{fig:zoom_amp}, \ref{fig:zoom_phase}, top and bottom, respectively). This zooming-in can help us to identify features that otherwise would have been missed, due to the short dynamical time-scales of Regimes 2 and 3, as discussed in the main text.

\section{Vertical spectrograms and resonances} \label{sec:appendix_vert_spec}
\begin{figure*}
	\includegraphics[width=15cm]{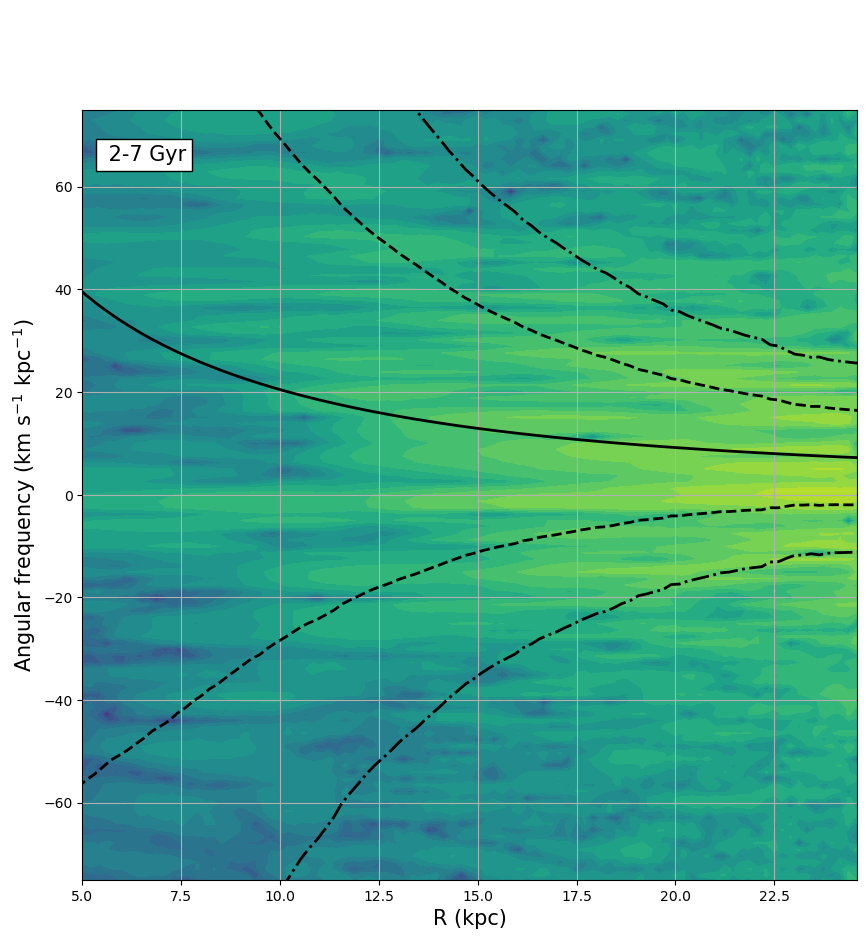}
    \caption{Same as Figure \ref{fig:vertspect_m1}, but showing the vertical spectrogram for the m=1 term  over a time baseline of 5 Gyr (from 2 and 7 Gyr in the simulation). Solid, dashed and dotted curves are as in Figure \ref{fig:vertspect_m1}. } 
    \label{fig:vertspect_m1_5Gyr}
\end{figure*}

The vertical spectrograms in Figure \ref{fig:vertspect_m1} and \ref{fig:vertspect_m2} are constructed over a temporal baseline of 1 Gyr. Such temporal baseline has been chosen to identify different phases, i.e. quiescent or with one (or more) interaction(s) with Sgr, to highlight the differences in the behaviour of the Galactic disc. However, 
a temporal baseline of 1 Gyr implies a resolution of 6.28 km/s/kpc on the angular frequency (y-axis of the vertical spectrogram) (see Equation 13 in \cite{2018MNRAS.480.4244C}), which is quite large. Figure \ref{fig:vertspect_m1_5Gyr} shows, as a complementary test, the vertical spectrogram over a baseline of 5 Gyrs, which implies a resolution of 1.26 km/s/kpc. 

\begin{figure*}
	\includegraphics[width=5cm]{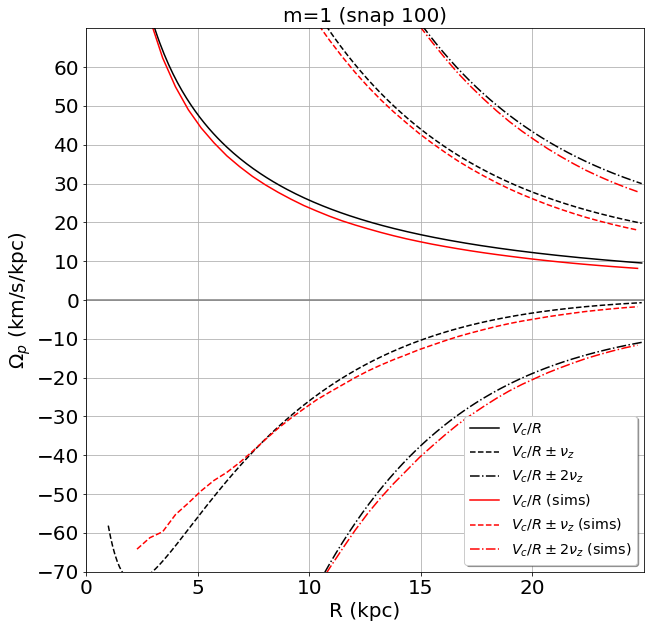}
	\includegraphics[width=5cm]{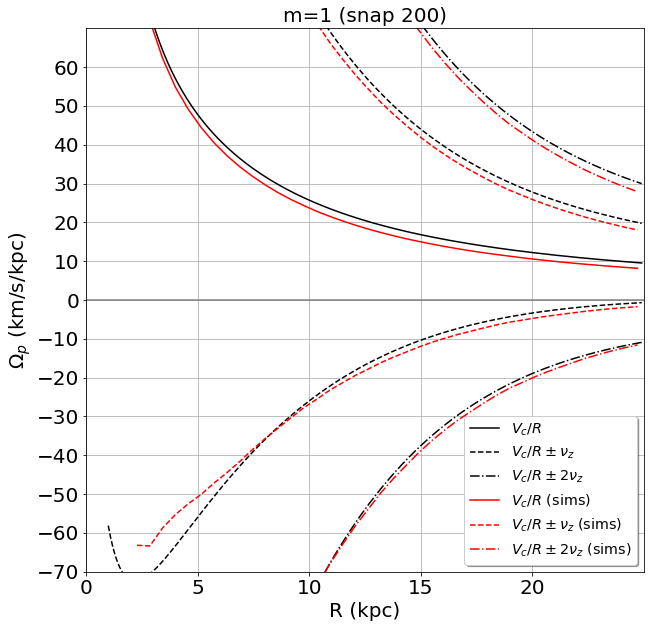}
	\includegraphics[width=5cm]{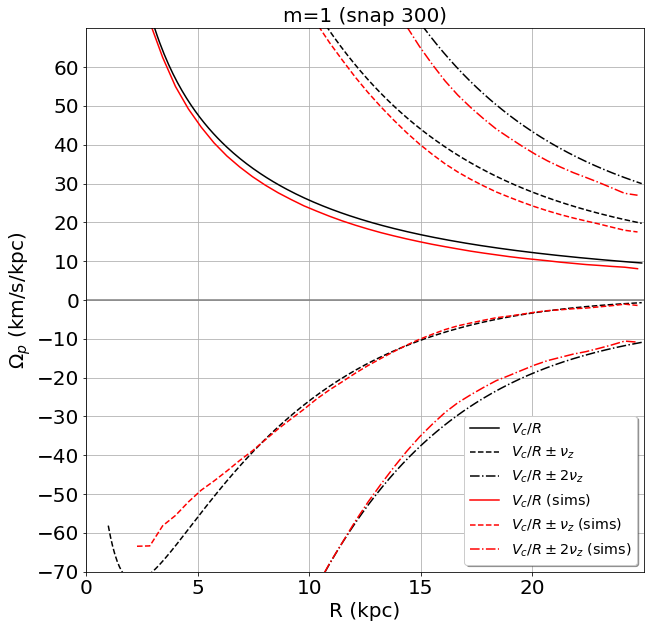}
	\includegraphics[width=5cm]{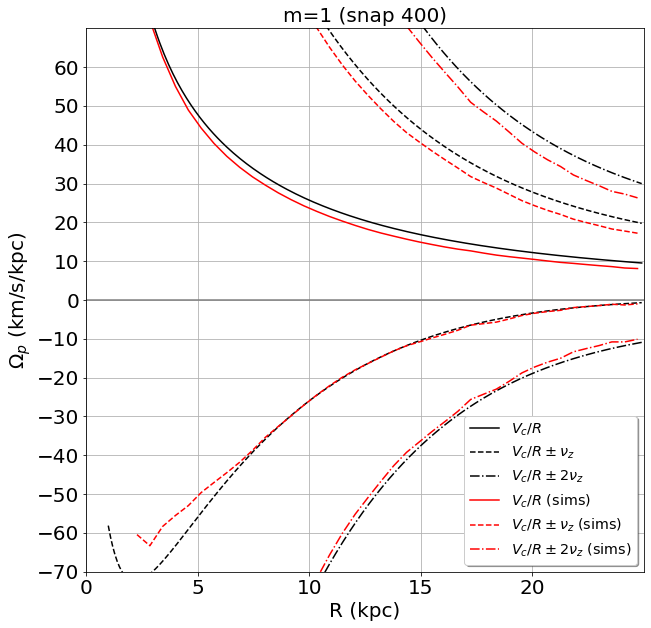}
	\includegraphics[width=5cm]{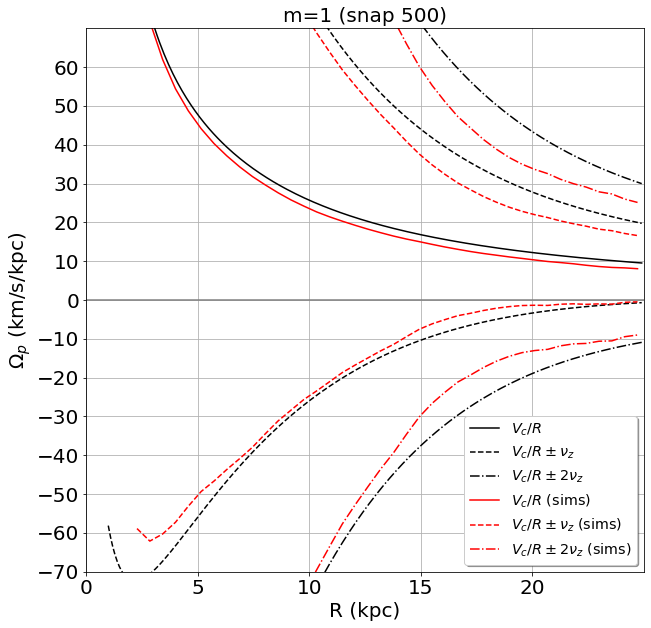}
	\includegraphics[width=5cm]{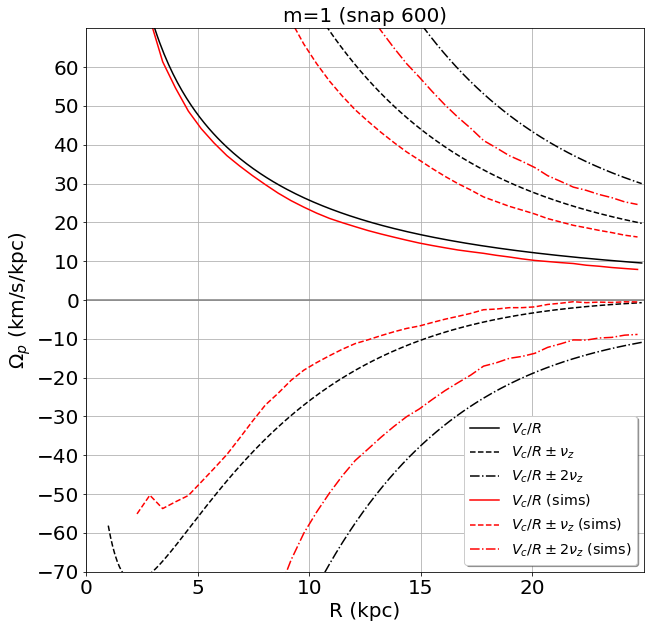}
    \caption{Comparison between vertical resonances from the analytical approximation (black curves) and those obtained using vertical frequencies from AGAMA at different times (red curves). (Top panels, from left to right: t=1,2,3 Gyr. Bottom panels, from left to right: 4,5,6 Gyr.) Solid lines show corotation, while the dashed and dashed-dotted curves show, respectively, $\Omega \pm \nu_z$ and $\Omega \pm 2 \nu_z$ vertical resonances.  } 
    \label{fig:spectrogram_curves}
\end{figure*}

Figure \ref{fig:spectrogram_curves} shows vertical resonances calculated in two different ways. Red curves are based on vertical frequencies obtained from the AGAMA package \citep{Vasilev2018,Vasilev2019}. The potentials for particles in each component in each snapshot are approximated using a spline interpolation of radial bins combined with fourier series (for disk components) and spherical harmonics (for spherical components). The actions and frequencies are then calculated using the Stackel fudge \citep{Binney2012,Sanders2015,Sanders2016}.

Black curves show the outcome of an analytical approximation, where vertical frequencies are written as the sum of the contribution from the halo, the bulge and the disc $\nu_z = \sqrt{ \nu^2_{z,\rm{halo}} + \nu^2_{z,\rm{bulge}} + \nu^2_{z,\rm{disc}} }$. Vertical frequencies from the halo and the bulge are calculated approximating them as Hernquist profiles, evaluating Eq. 5 of \cite{Donghia16b} at z=0. Following \cite{Donghia2013}, we describe the disc density distribution as
\begin{equation}
  \rho(R,z) = \frac{ M_* }{ 4 \pi z_0 R^2_s} \rm{sech}^2 \frac{z}{z_0} exp(-R/R_s)) \quad ,
\end{equation}
where $R_s$ is the scale length of the disc, $z_0$ is the scale height of the disc, and $M_*$ is the mass of the disc. Using the above density distribution, the vertical frequency $\nu^2_{z,\rm{disc}}$ can be approximated as
\begin{equation}
  \nu^2_{z,\rm{disc}} =  \frac{ G M_* }{ z_0 R^2_s} \exp(-R/R_s) \quad.
\end{equation}
Since the analytical approximation reflects the initial conditions, the agreement between the black and red curves is better at the beginning of the simulation (upper left panel) and becomes worse at the end of the simulation (lower right panel).

\bsp	
\label{lastpage}
\end{document}